\documentclass[12pt,a4paper]{article}
\usepackage{amsmath,amssymb,graphicx,epsfig,cite,color}
\usepackage[left=2.7cm,right=2.7cm,top=2.5cm,bottom=2.5cm]{geometry}
\usepackage{subfigure}
\usepackage[table]{xcolor}
\usepackage{booktabs}

\usepackage{graphicx}
\usepackage{lscape}
\usepackage{appendix}
\usepackage{multirow}
\usepackage[colorlinks=true,linkcolor=red]{hyperref}%

\usepackage{longtable}

\begin{document}
    \title{\textbf{Surface Terms of Quintic Quasitopological Gravity and Thermodynamics of Quasi-Topological Magnetic Brane Coupled to Nonlinear Electrodynamics}}

\author{\vspace{1cm}\\
    {\small
        A.~Bazrafshan$^{1}$,
        A. R. Olamaei$^{1,2}$} \\
    {\small $^1$ Department of Physics, Jahrom University, Jahrom, P.~ O.~ Box 74137-66171, Iran}\\
    {\small$^2$  School of Particles and Accelerators, Institute for Research in Fundamental Sciences (IPM),}\\
        {\small P. O. Box 19395-5531, Tehran, Iran }}

\date{}

\begin{titlepage}
    \maketitle
    \thispagestyle{empty}

\begin{abstract}
For the the quintic quasitopological action which has no well-defined variational principle, we introduced a surface term that for a spacetime with flat boundaries make the action well-defined. 
Moreover, we investigated the numerical solutions of the above-mentioned gravity coupled to the nonlinear logarithmic and exponential electrodynamics. It has no horizon and curvature except one conical singularity at $r=0$ with a deficit angle $\delta\phi$.
Also we found the counterterm which removes non-logarithmic divergences for the static quintic quasitopological gravity. Using this counterterm one can calculate a finite action and conserved quantities for the quintic quasitopological gravity.
\end{abstract}

\end{titlepage}


\section{Introduction}
The AdS/CFT correspondence or more generally the bulk/boundary or gauge/gravity duality provides a framework to study the nonperturbative regime of quantum field theory via holographic studies. This duality indicates that the Einstein gravity in the bulk corresponds to a strongly coupled conformal filed theory (CFT) with large number of colors $N_c$ which lives on the boundary. Accordingly, one can calculate a quantity from the weak gravity side and find its corresponding dual on the strongly correlated CFT or vise versa using the dictionary of duality  \cite{AdSCFT}.

For Einstein gravity with a two-derivative bulk action, the AdS/CFT conjecture can be used only in the limit of large $N_c$ and large $\lambda$. On the other hand, based on the gauge/gravity dictionary, the coupling constant in the gravity side find their dual as the central charges in the gauge side.
Thus, Einstein gravity is only dual to the limited class of CFTs with all central charges equal to each other and does not have enough parameters to account for the CFTs with different central charges.
Considering the case with finite $\lambda$ and $N_c$, which needs the central charges to be different, has to have its correspondent dual gravity with enough parameters to account for the ratios between central charges.
Therefore, to extend the duality conjecture to conformal field theories with different central charges, higher order gravities, which are rich enough to contain parameters necessary to cope with different dual central charges, has to be considered. \cite{Hendi22,Tangher,Hassaine:2007py,Bek1,Mart1,Henn,Mart2}.
In dimensions more than 4, it is possible to add terms to the Lagrangian such that it satisfies Einstein's assumptions and yields to the second-order equations of motion. To this end, various theories have been proposed considering gravity in higher dimensions.
One modification is to add the Gauss-Bonnet term to the gravity action which adds just one more coupling constant to the theory where its dual CFTs has limited range.
Another kind of modified theories is the Lovelock gravity \cite{Love,Myers1,Hofman,Friedman1,Friedman2,Jacobson}. It is the most general theory which is linear in the second derivatives of the metric and has no higher metric derivatives in all dimensions. Although the equations of motions yield to the second-order differential equations for the Lovelock theory of order $k$, but the $k$'th order terms in this theory will not appear until for the dimensions $n>2k$ and do not take any part in lower dimensions.
Therefore, a new toy model with cubic curvature terms in the Lagrangian has been proposed which contributes in five dimensions and for the spherically symmetric metric, the resulting equations of motion are second-order differential equation. Some interesting higher derivative theories of gravity in arbitrary dimensions are introduced in Refs. \cite{Oliva1,Oliva2,Oliva3,Myers:2010ru}. In six dimensions, similar to the third order Lovelock gravity, the cubic term does not contribute.
Although the topological origin of Euler density in six dimensions causes this property of Lovelock gravity, it is not the case for the new theory \cite{Myers:2010ru} and therefore is called quasitopological gravity.

In general, higher order curvature theories like quasitopological gravity participate in field equations except for $n=2k-1$, where $k$ is the curvature order of the theory. There are many studies in this topic including cubic and quartic quasitopological gravity \cite {Lemos1,Lemos2,Lemos3,Aminneborg,Mann,Cai,Dehghani1,Dehghani2,Dehghani3,Brenna2012,Ghanaatian1,Ghanaatian3}. Also the explicit Lagrangian for the quintic quasitopological gravity is given in \cite{Quintic1}.

There are several advantages of studying quintic quasitopological gravity which motivates this work. First of all, its asymptotically AdS solution has a broader class of dual CFTs with respect to lower quartic and cubic forms. Second, the requirement of positive energy fluxes \cite{Brig,Ge,Boer} causes five constraints that leads to the specification of five coupling constants. Third, against the cubic and quartic quasitopological gravities which are unique, the uniqueness of the quintic form is under question. 
Originally, a quintic quasitopological theory in five dimensions was presented in \cite{Cisterna:2017umf} and recently, by solving a recurrence equation, a systematic method to construct quasitopological Lagrangians of any order introduced in \cite{Bueno:2019ycr}.

It should be noted that, like the Hilbert-Einstein (H-E) action, the variation of the quasitopological action is not well-defined. It's because we face a derivative of the variation of the metric normal to the boundary in a surface integral. It can be removed by introducing a surface term to the action \cite{Gib}.
The necessity of adding the surface term have been emphasized in both the path integral formalism of quantum gravity \cite{Hawking:1979ig} and Hamiltonian framework of general relativity \cite{York72}.
So there is a strong motivation to study the surface terms in higher curvature gravity such as Lovelock or quasitopological gravity.
The surface term corresponding to the Lovelock and quasitopological gravity of the third and fourth order are introduced in Refs. \cite{SurLove,DBS,Ghanaatian2} respectively. Our aim here is to introduce the surface term for the quasitopological gravity of the fifth order in $(n+1)$ dimensions in order to solve the above-mentioned problem as well as providing it for the Hamiltonian formalism.

As in the case of Einstein gravity, as the boundary goes to infinity, the action and conserved quantities diverge. Therefore, to calculate the conserved quantities of the solutions, we introduce a counterterm to regulate the theory and remove the divergences.

Another goal of this work is to calculate the black brane solutions in the presence of two kinds of nonlinear electrodynamics terms of the Born-Infeld (BI) kind.  One of the motivations for introducing nonlinear electrodynamics is to remove the infinite self-energy of the point-like-charges. The second one is its compatibility with string theory and AdS/CFT correspondence, and the third is that the pair creation in Hawking radiation can be described by that \cite{Aros,Chen,Fuk}. There are also several applications of nonlinear electrodynamics theory from cosmological models \cite{Dyad}, description of the inflationary epoch and the late-time accelerated expansion of the universe \cite{Nove}, to finding the first exact regular black hole solutions with a nonlinear electrodynamics source which satisfies the weak energy condition \cite{Ayon}.
The first type of nonlinear electrodynamics was introduced by Born and Infeld \cite{BI}. The importance of this theory is that it emerges as the low-energy limit of the open-string theory and can be used in the D-branes and AdS/CFT duality descriptions \cite{Fradkin}. In recent years, different versions of nonlinear electrodynamics are proposed among them one can name logarithmic \cite{Soleng} and exponential Lagrangians \cite{Hendi1}.
They reduce to the linear Maxwell Lagrangian as the nonlinear parameter goes to infinity. The divergence of the electric field at the origin can be removed by the logarithmic form  \cite{Sheykhi1} as in the case of Born-Infeld, whereas the exponential form just transforms the singularity into a weaker type than that of Einstein-Maxwell theory\cite{Sheykhi2}.

Some kinds of black hole and black brane solutions are studied in quasitopological gravity \cite{Ghanaa2,Ghanaa3,Ghanaa4,Ghanaa5}. One can find the thermodynamic properties of rotating black holes and black brane in the quartic quasitopological Lifshitz gravity in the presence of nonlinear BI field in \cite{Ghana1}. 
Since magnetic branes solutions have conical geometry and are horizonless, they attracted much interest and are investigated thoroughly in recent years. The magnetic brane solutions in the presence of BI field in the cubic and quartic quasitopological gravity are investigated in \cite{Taghi,Ghanaa6} respectively.
Moreover, excluding the location of line source, magnetic brane solutions are flat everywhere.
Here we study the thermodynamic properties of the black brane solutions in the quintic quasitopological gravity in the presence of logarithmic and exponential nonlinear electromagnetic field.

The outline of the paper is as follows. In the next section, \ref{Bound}, we take a brief look at the quintic quasitopological gravity and also review the surface terms which make the Einstein, Gauss-Bonnet, cubic and quartic quasitopological gravity actions well defined. Then we introduce the surface term for the quintic quasitopological gravity for flat spacetime and plot the numerical solutions of the relevant quintic equations as well as analyzing their various properties. Also we will show that these solutions have no horizon but a conical singularity with the deficit angle $\delta\phi$.
Next, we generalize our solutions to
spinning cases in arbitrary
dimensions. In section \ref{Numerics} we calculate the physical properties of the magnetic branes like mass and charge. This will be done by introducing the counterterm for the quintic quasitopological gravity which makes the action and conserved quantities finite. In sections \ref{spin} and \ref{struc} we generalize the static spacetime to rotating solutions and give the physical properties of magnetic brane respectively.
Last section, \ref{result}, is devoted to closing remarks.

\section{\ Surface Terms of Quintic Quasitopological Gravity \label{Bound}}
The $(n+1)$ dimensional action for the quintic quasitopological gravity in the presence of nonlinear electrodynamics is written as
\begin{equation}\label{action1}
I_{G}=\frac{1}{16\pi}\int{d^{n+1}x\sqrt{-g}\big\{-2\Lambda+{\mathcal L}_1+\mu_{2}{\mathcal L}_2+\mu_{3}{\mathcal L}_3+\mu_{4}{\mathcal L}_4+\mu_{5}{\mathcal L}_5+\mathcal{L}(F)\big\}},
\end{equation}
where $\Lambda=-\frac{n(n-1)}{2l^2}$ is the cosmological constant for the anti de-Sitter space. $\mathcal{L}_1=R$ is the H-E and ${\mathcal L}_2=R_{abcd}R^{abcd}-4R_{ab}R^{ab}+R^2$ the quadratic Lovelock (Gauss-Bonnet) Lagrangian. Also $\mathcal{L}_3$, $\mathcal{L}_4$ and $\mathcal{L}_5$ are the cubic, quartic and quintic quasitopological Lagrangians respectively with following definitions:
\begin{eqnarray}
{{\mathcal L}_3}&=&
R_a{{}^c{{}_b{{}^d}}}R_c{{}^e{{}_d{{}^f}}}R_e{{}^a{{}_f{{}^b}}}+\frac{1}{(2n-1)(n-3)} \bigg(\frac{3(3n-5)}{8}R_{abcd}R^{abcd}R-3(n-1)R_{abcd}R^{abc}{{}_e}R^{de}\nonumber\\
&&+3(n+1)R_{abcd}R^{ac}R^{bd}+6(n-1)R_a{{}^b}R_b{{}^c}R_{c}{{}^a}-\frac{3(3n-1)}{2}R_a{{}^b}R_b{{}^a}R \nonumber \\
 &&+\frac{3(n+1)}{8}R^3\bigg),
\end{eqnarray}
\begin{eqnarray}\label{quartic}
{\mathcal{L}_4}&=& b_{1}R_{abcd}R^{cdef}R^{hg}{{}_{ef}}R_{hg}{{}^{ab}}+b_{2}R_{abcd}R^{abcd}R_{ef}{{R}^{ef}}+b_{3}RR_{ab}R^{ac}R_c{{}^b}+b_{4}(R_{abcd}R^{abcd})^2\nonumber\\
&&+b_{5}R_{ab}R^{ac}R_{cd}R^{db}+b_{6}RR_{abcd}R^{ac}R^{db}+b_{7}R_{abcd}R^{ac}R^{be}R^d{{}_e}+b_{8}R_{abcd}R^{acef}R^b{{}_e}R^d{{}_f}\nonumber\\
&&+b_{9}R_{abcd}R^{ac}R_{ef}R^{bedf}+b_{10}R^4+b_{11}R^2 R_{abcd}R^{abcd}+b_{12}R^2 R_{ab}R^{ab}\nonumber\\
&&+b_{13}R_{abcd}R^{abef}R_{ef}{{}^c{{}_g}}R^{dg}+b_{14}R_{abcd}R^{aecf}R_{gehf}R^{gbhd},
\end{eqnarray}
and
\begin{eqnarray}\label{quintic}
{\mathcal{L}_5}&=&
c_{1} R R_{b}^{a} R_{c}^{b} R_{d}^{c} R_{a}^{d}+c_{2} R R_{b}^{a} R_{a}^{b} R_{ef}^{cd} R_{cd}^{ef}+c_{3} R R_{c}^{a} R_{d}^{b} R_{ef}^{cd} R_{ab}^{ef}+c_{4} R_{b}^{a} R_{a}^{b} R_{d}^{c}  R_{e}^{d} R_{c}^{e}\nonumber\\
&&+c_{5} R_{b}^{a} R_{c}^{b} R_{a}^{c}  R_{fg}^{de} R_{de}^{fg}+c_{6} R_{b}^{a} R_{d}^{b} R_{f}^{c}  R_{ag}^{de} R_{ce}^{fg}+c_{7} R_{b}^{a} R_{d}^{b} R_{f}^{c} R_{cg}^{de} R_{ae}^{fg}+c_{8} R_{b}^{a} R_{c}^{b} R_{ae}^{cd} R_{gh}^{ef} R_{df}^{gh}\nonumber\\
&&+c_{9} R_{b}^{a} R_{c}^{b} R_{ef}^{cd} R_{gh}^{ef} R_{ad}^{gh}+c_{10} R_{b}^{a} R_{c}^{b} R_{eg}^{cd} R_{ah}^{ef} R_{df}^{gh}+c_{11} R_{c}^{a} R_{d}^{b} R_{ab}^{cd} R_{gh}^{ef} R_{ef}^{gh}+c_{12} R_{c}^{a} R_{d}^{b} R_{ae}^{cd} R_{gh}^{ef} R_{bf}^{gh}\nonumber\\
&&+c_{13} R_{c}^{a} R_{d}^{b} R_{ef}^{cd} R_{gh}^{ef} R_{ab}^{gh}+c_{14} R_{c}^{a} R_{d}^{b} R_{eg}^{cd} R_{ah}^{ef} R_{bf}^{gh}+c_{15} R_{c}^{a} R_{e}^{b} R_{af}^{cd} R_{gh}^{ef} R_{bd}^{gh}+c_{16} R_{b}^{a} R_{ad}^{bc} R_{fh}^{de} R_{ci}^{fg} R_{eg}^{hi}\nonumber\\
&&+c_{17} R_{b}^{a} R_{de}^{bc} R_{cf}^{de} R_{hi}^{fg} R_{ag}^{hi}+c_{18} R_{b}^{a} R_{df}^{bc} R_{ac}^{de} R_{hi}^{fg} R_{eg}^{hi}+c_{19} R_{b}^{a} R_{df}^{bc} R_{ah}^{de} R_{ei}^{fg} R_{cg}^{hi}+c_{20} R_{b}^{a} R_{df}^{bc} R_{gh}^{de} R_{ei}^{fg} R_{ac}^{hi}\nonumber\\
&&+c_{21} R_{cd}^{ab} R_{eg}^{cd} R_{ai}^{ef} R_{fj}^{gh}R_{bh}^{ij}+c_{22} R_{ce}^{ab} R_{af}^{cd} R_{gi}^{ef} R_{bj}^{gh}R_{dh}^{ij}+c_{23} R_{ce}^{ab} R_{ag}^{cd} R_{bi}^{ef} R_{fj}^{gh}R_{dh}^{ij}\\&&+c_{24} R_{ce}^{ab} R_{fg}^{cd} R_{hi}^{ef} R_{aj}^{gh}R_{bd}^{ij}\nonumber.
\end{eqnarray}
The coefficients $b_i$ for the quartic and $c_i$ for the quintic quasitopological Lagrangians are numerous and lengthy and introduced in Tables \ref{tab:quart} and \ref{tab:quint} of the Appendix respectively. They are obtained considering the constraint that the equations of motion have to be second order in metric derivatives. For $c_i$, this leads to many linear equations which some of them are not independent. Reducing them to the system of linear independent equations, leaves us one degree of freedom which leads to infinite class of answers for the coefficients $c_i$. In Table \ref{tab:quint} we just present one possible class of answers.

${\cal L}(F)$ is nonlinear electrodynamics Lagrangian with the following definition:
\begin{eqnarray}
\mathcal{L}(F)=\left\{
\begin{array}{ll}
$$4\beta^2[\mathrm{exp}(-\frac{F}{4\beta^2})-1]$$,\quad\quad\quad \quad\quad  \ {EN}\quad &  \\ \\
$$-8\beta^2 \mathrm{ln}[1+\frac{F}{8\beta^2}]$$.\quad\quad\quad\quad\quad\quad  \ {LN}\quad &
\end{array}
\right.
\end{eqnarray}
$\beta$ is the nonlinearity parameter and $F=F_{\mu\nu}F^{\mu\nu}$, where $F_{\mu\nu}$ is the electromagnetic field tensor. It is determined as $F_{\mu\nu}=\partial_{\mu}A_{\nu}-\partial_{\nu}A_{\mu}$ where $A_{\mu}$ is the four-vector potential. In the weak field approximation, which is characterized by $\beta\rightarrow\infty$, the nonlinear theory approaches the linear Maxwell ${\cal}L(F)=-F_{\mu\nu}F^{\mu\nu}$.
In the following we use the dimensionless coefficients as

\begin{eqnarray}
\mu_{3}=\frac{8(2n-1)\hat{\mu}_{3}l^4}{(n-2)(n-5)(3n^2-9n+4)},
\end{eqnarray}
\begin{eqnarray}
\mu_{4}=\frac{\hat{\mu}_{4}l^6}{n(n-1)(n-3)(n-7)(n-2)^2(n^5-15n^4+72n^3-156n^2+150n-42)},
\end{eqnarray}
\begin{eqnarray}\label{mu5}
\mu_{5}&=&\frac{\hat{\mu}_{{5}}{l}^{8}}{(n-3)(n-9)(n-2)^2 } (8\,{n}^{12}+26\,{n}^{11}-1489\,{n}^{10}+11130\,{n}^{9}-26362\,{n}^{8} \nonumber \\ &&-
75132\,{n}^{7}
+705657\,{n}^{6}-2318456\,{n}^{5}+4461054\,{n}^{4}-
5484168\,{n}^{3} \nonumber\\
&&+4290516\,{n}^{2}-1968224\,n+405376)^{-1}~.
\end{eqnarray}
In general, when the manifold has boundaries, the variation of the action with respect to the metric is not well defined since one encounters a total
derivative that produces a surface integral involving the derivatives of $%
\delta g_{\mu \nu }$ normal to the boundary $\partial \mathcal{M}$. These
normal derivatives of $\delta g_{\mu \nu }$ can be canceled by the variation
of the surface action \cite{bostani}:

\begin{equation*}
I_{surface}=\frac{1}{8\pi }\int_{\partial \mathcal{M}}d^{n+1}x\sqrt{-\gamma }%
\sum_{p=1}^{\left[ n/2\right] }\alpha _{p}Q_{p},
\end{equation*}%
where
\begin{eqnarray}
Q_{p} &=&p\int\limits_{0}^{1}dt\,\delta _{\lbrack ii_{1}\cdots
    i_{2p-1}]}^{[jj_{1}\cdots j_{2p-1}]}\,K_{j_{1}}^{i_{1}}\,\times  \notag \\
&&\times \left( \frac{1}{2}\,\hat{R}_{j_{2}j_{3}}^{i_{2}i_{3}}(\gamma
)-t^{2}K_{j_{2}}^{i_{2}}K_{j_{3}}^{i_{3}}\right) \cdots \left( \frac{1}{2}\,%
\hat{R}_{j_{2p-2}j_{2p-1}}^{i_{2p-2}i_{2p-1}}(\gamma
)-t^{2}\,K_{j_{2p-2}}^{i_{2p-2}}K_{j_{2p-1}}^{i_{2p-1}}\right)~,
\label{SurLag}
\end{eqnarray}%
where $\gamma _{ab}$ and $K_{ab}=-\gamma _{a}^{\mu }\nabla
_{\mu }n_{b}$ are the induced metric and extrinsic curvature of the boundary
$\partial \mathcal{M}$ respectively, and $n_b $ is the timelike unit four-vector normal to the boundary.
They are constructed in a sector of the theory which leads to second order field equations.
The surface term of the H-E action that makes it well defined is

\begin{eqnarray}
I_{b}^{(1)}=\frac{1}{8\pi}\int_{\partial\mathcal{M}} d^{n}x \sqrt{-\gamma}K.
\end{eqnarray}
 As we are interested in studying the space-times with flat boundary, $\hat{R}_{abcd}(\gamma)=0$. Therefore the surface term corresponding to the quadratic Lovelock  is
\begin{equation}
I_{b}^{(2)}=\frac{1}{8\pi }\int_{\partial \mathcal{M}}d^{n}x\sqrt{-\gamma }%
\left\{ \frac{2\hat{\mu _{2}}l^{2}}{(n-2)(n-3)}J\right\} ,  \label{Ib2}
\end{equation}%
where $J$ is the trace of
\begin{equation}
J_{ab}=\frac{1}{3}%
(2KK_{ac}K_{b}^{c}+K_{cd}K^{cd}K_{ab}-2K_{ac}K^{cd}K_{db}-K^{2}K_{ab}).
\label{Jab}
\end{equation}%
The surface terms for the curvature-cubed term of quasitopological gravity
have been introduced in \cite{DV} as
\begin{eqnarray}
&&I_{b}^{(3)}=\frac{1}{8\pi }\int_{\partial \mathcal{M}}d^{n}x\sqrt{-\gamma }%
\Big\{\frac{3\hat{\mu _{3}}l^{4}}{5n(n-2)(n-1)^{2}(n-5)}%
(nK^{5}-2K^{3}K_{ab}K^{ab}  \notag \\
&&\,\ \ \ \ \ \ \ \ \ \ \ \ \ \ \ \ \ \ \ \ \ \ \ \ \ \ \ \
+4(n-1)K_{ab}K^{ab}K_{cd}K_{e}^{d}K^{ec}-  \notag \\
&&\,\ \ \ \ \ \ \ \ \ \ \ \ \ \ \ \ \ \ \ \ \ \ \ \ \ \ \ \
(5n-6)KK_{ab}[nK^{ab}K^{cd}K_{cd}-(n-1)K^{ac}K^{bd}K_{cd}])\Big\},
\label{Ib3}
\end{eqnarray}
and for quartic one it is
\begin{eqnarray}
&&I_{b}^{(4)}={\frac{1}{8\pi }}\int_{\partial \mathcal{M}}d^{n}x\sqrt{%
    -\gamma }{\frac{\hat{2\mu _{4}}{l}^{6}}{7n(n-1)\left( n-2\right) \left(
        n-7\right) \left( {n}^{2}-3\,n+3\right) }}\Big\{\alpha
_{1}K^{3}K^{ab}K_{ac}K_{bd}K^{cd}  \notag \\
&&\,\ \ \ \ \ \ \ \ \ \ \ \ \ \ \ \ \ \ \ \ \ \ \ \ \ \ \ \ +\alpha
_{2}K^{2}K^{ab}K_{ab}K^{cd}K_{c}^{e}K_{de}+\alpha
_{3}K^{2}K^{ab}K_{ac}K_{bd}K^{ce}K_{e}^{d}  \notag \\
&&\,\ \ \ \ \ \ \ \ \ \ \ \ \ \ \ \ \ \ \ \ \ \ \ \ \ \ \ \ +\alpha
_{4}KK^{ab}K_{ab}K^{cd}K_{c}^{e}K_{d}^{f}K_{ef}+\alpha
_{5}KK^{ab}K_{a}^{c}K_{bc}K^{de}K_{d}^{f}K_{ef}   \\
&&\,\ \ \ \ \ \ \ \ \ \ \ \ \ \ \ \ \ \ \ \ \ \ \ \ \ \ \ \ +\alpha
_{6}KK^{ab}K_{ac}K_{bd}K^{ce}K^{df}K_{ef}+\alpha
_{7}K^{ab}K_{a}^{c}K_{bc}K^{de}K_{df}K_{eg}K^{fg}\Big\}\notag  \label{Ib4},
\end{eqnarray}%
where
\begin{eqnarray}
\alpha _{1} &=&1  \notag \\
\alpha _{2} &=&2\left( {n}^{3}-4\,{n}^{2}+3\,n+3\right),  \notag \\
\alpha _{3} &=&-6n(n-1)^{2},  \notag \\
\alpha _{4} &=&-4n\left( {n}^{2}-3\,n+3\right) (n+2),  \notag \\
\alpha _{5} &=&-\,2n\left( n-3\right) \left( {n}^{2}-n-3\right),  \notag \\
\alpha _{6} &=&2(n-1)\left( 3\,{n}^{3}-{n}^{2}-9\,n+12\right),  \notag \\
\alpha _{7} &=&24.  \label{alph}
\end{eqnarray}%

In this work, to make the quintic quasitopological action well defined, we introduce its corresponding surface term follows:

\begin{equation}
I_{b}^{(5)}={\frac{1}{8\pi }}\int_{\partial \mathcal{M}}d^{n}x\sqrt{%
    -\gamma }{\hat{\mu _{5}}l^8}(\alpha'_{1}K^{2}K^{j}_{s}K^{m}_{n}K^{n}_{m}K^{p}_{q}K^{q}_{j}K^{w}_{p}K^{s}_{w} $$$$
+\alpha'_{2}K^{2}K^{j}_{s}K^{m}_{n}K^{n}_{j}K^{p}_{m}K^{q}_{p}K^{w}_{q}K^{s}_{w}+\alpha'_{3}K^{2}K^{j}_{s}K^{m}_{n}K^{n}_{p}K^{p}_{m}K^{q}_{w}K^{w}_{j}K^{s}_{q} $$$$
 +\alpha'_{4}KK^{h}_{j}K^{j}_{s}K^{m}_{n}K^{n}_{m}K^{p}_{q}K^{q}_{h}K^{w}_{p}K^{s}_{w}+\alpha'_{5}KK^{h}_{j}K^{j}_{w}K^{m}_{n}K^{n}_{m}K^{p}_{q}K^{q}_{p}K^{w}_{s}K^{s}_{h} $$$$
 +\alpha'_{6}KK^{h}_{j}K^{j}_{s}K^{m}_{n}K^{n}_{p}K^{p}_{m}K^{q}_{w}K^{w}_{h}K^{s}_{q}+\alpha'_{7}KK^{h}_{j}K^{j}_{s}K^{m}_{n}K^{n}_{h}K^{p}_{m}K^{q}_{p}K^{w}_{q}K^{s}_{w}$$$$
 +\alpha'_{8}Kk^{h}_{j}K^{j}_{w}K^{m}_{n}K^{n}_{m}K^{p}_{q}K^{q}_{p}K^{w}_{s}K^{s}_{h}+\alpha'_{9}K^{h}_{j}k^{j}_{w}K^{w}_{h}K^{m}_{n}K^{n}_{p}K^{p}_{m}K^{q}_{h}K^{h}_{i}K^{i}_{q}) \label{Ib5}.
\end{equation}
Using the metric
\begin{eqnarray}\label{metr}
ds^2=-\frac{\rho^2}{l^2}dt^2+\frac{d\rho^2}{f(\rho)}+l^2 g(\rho)d \phi^2+\frac{\rho^2}{l^2}dX^2,
\end{eqnarray}
where $dX^2=\sum_{i=1}^{n-2} dx_i^2$ is an $(n-2)$-dimensional hypersurface with the form of Euclidean metric in the volume $V_{n-2}$. $\rho$ and $\phi$ are the radial and angular coordinates respectively where $\phi$ is dimensionless and has the range $0\leq \phi <2\pi$.\\
Requiring that after varying $I_{b}^{(5)}$, all the normal derivatives $\delta g_{\mu\nu}$ must vanish, one can find the $\alpha_i$ coefficients as:

\begin{eqnarray}
\alpha'_{1} &=&\frac {40(n^4-4n^3+3n^2+2n+2)}{9n\left( -72+188n-182n^2+81n^3-16n^4+n^5\right)(n-1) },  \notag \\
\alpha'_{2} &=&-\frac {80(n^4-3n^3+n^2+n+1)}{9n^2
        ( -72+188n-182n^2+81n^3-16n^4+n^5) },
  \notag \\
\alpha'_{3} &=&\frac {40(n^4-4n^3+5n^2-2n-2)}{9n
        ( -72+188n-182n^2+81n^3-16n^4+n^5) (n-1) },
  \notag \\
\alpha'_{4} &=&{\frac {80({n}^{3}-2{n}^{2}-n-1)}{9{n}^{2} ( -72+
        188n-182{n}^{2}+81{n}^{3}-16{n}^{4}+{n}^{5} ) }},
  \notag \\
\alpha'_{5} &=&-{\frac {10(8-20{n}^{2}-4{n}^{3}+13{n}^{4}-6{n}^
        {5}+{n}^{6})}{ 9(n-1) {n}^{2}( -72+188n-182{n}^{2}+81{n}^{3}-16{n}^{4}+{n}^{5}) }},
  \notag \\
\alpha'_{6} &=&-{\frac {40(4-4{n}^{2}+4{n}^{3}+3{n}^{4}-4{n}^{5
        }+{n}^{6})}{9( n-1) {n}^{2} (-72+188n-182{n}^{2}+
        81{n}^{3}-16{n}^{4}+{n}^{5}) }},
  \notag \\
\alpha'_{7} &=&\frac {10(5{n}^{2}+5n-2)}{9n( {n}^{3}-12{n}^{
            2}+29n-18)},
 \notag \\
\alpha'_{8} &=&-{\frac { 20( n+1) ^{2}}{9n( {n}^{4}-13{n}^{3}+41{n}^{2}-47n+18 ) }},
 \notag \\
\alpha'_{9} &=&{\frac {80}{9{n}^{2} (72-260n+370{n}^{2}-263{n}^{3}+97{n}^{4}-17{n}^{5}+{n}^{6}) }}.
\label{alph1}
\end{eqnarray}

Therefore, if we consider $I_b$ as $I_{b}^{(1)} + \cdots + I_{b}^{(5)}$, the quasitopological action would be well defined.

\section{Numerical Analysis}\label{Numerics}
Now we concentrate on the numerical solutions of the quintic quasitopological gravity having into account the presence of nonlinear electromagnetic field.
Defining $\Psi= -\frac{l^2}{\rho^2}f(\rho)$, and integrating the action (\ref{action1}) by parts yields
\begin{eqnarray}\label{Act2}
S&=&\frac{n-1}{16\pi l^2}\int d^{n} x\int d\rho N(\rho)\nonumber\\&&\times\bigg\{\bigg[\rho^n\bigg(1+\Psi+\hat{\mu}_2\Psi^2+\hat{\mu}_3\Psi^3+\hat{\mu}_4\Psi^4+\hat{\mu}_5\Psi^5\bigg)\bigg]^{'}\nonumber\\&&+
    \left\{
    \begin{array}{ll}
    $$\frac{4\beta^2 l^2 \rho^{n-1}}{n-1}\bigg[\mathrm{exp}\bigg(-\frac{h^{'2}}{2l^2\beta^2 N^2(\rho)}\bigg)-1\bigg]\bigg\}$$,\quad  \ {EN}\quad &  \\ \\
    $$-\frac{8\beta^2 l^2 \rho^{n-1}}{n-1}\mathrm {ln}\bigg[1+\frac{h^{'2}}{4\beta^2 l^2N^2(\rho)}\bigg]\bigg\}$$,\quad\quad\quad  \ {LN}\quad &
    \end{array}
    \right.
\end{eqnarray}
where the definitions $A_\mu = h(\rho)\delta^{\phi}_{\mu}$ for the vector potential and $g(\rho)=N^2(\rho)f(\rho)$ are employed.
Variation of the action with respect to $\Psi(\rho)$ yields to

\begin{eqnarray}
\big(1 + 2\hat{\mu}_{2}\Psi + 3\hat{\mu}_{3}\Psi^2 + 4\hat{\mu}_{4}\Psi^3 + 5\hat{\mu}_{5}\Psi^5\big)\frac{dN(\rho)}{d\rho} = 0.
\end{eqnarray}
For the above equation to be satisfied for all values of $\Psi$, it is obvious that $N(\rho)$ has to be constant and we take it to be one.
Varying the action with respect to $h(\rho)$ we get to the following equations:

\begin{eqnarray}\label{equ2}
\bigg\{(n-1)\rho^n\bigg(1+\Psi+\hat{\mu}_{2} \Psi^2+\hat{\mu}_{3} \Psi^3+\hat{\mu}_{4}\Psi^4+\hat{\mu}_{5}\Psi^5\bigg)\bigg\}^{'} \\ +\left\{
\begin{array}{ll}
$$4\rho^{n-1}(l^2 \beta^2+h^{'2})\mathrm {exp}\bigg(-\frac{h^{'2}}{2 l^2\beta^2}\bigg)\nonumber\\-4l^2\beta^2\rho^{n-1}=0$$,\quad\quad\quad\quad \quad\quad\quad  \ {EN}\quad &  \\ \\
$$-8\beta^2 l^2 \rho^{n-1} \mathrm{ln} (1+\frac{h^{'2}}{4\beta^2 l^2})\nonumber\\+4\rho^{n-1}h^{'2}(1+\frac{h^{'2}}{4\beta^2 l^2})^{-1}=0$$,\quad\quad  \ {LN}\quad &
\end{array}
\right.
\end{eqnarray}
and
\begin{equation}\label{equ3}
\left\{
\begin{array}{ll}
$$\bigg(\rho^{n-1}h^{'}\mathrm {exp}\bigg[-\frac{h^{'2}}{2l^2\beta^2}\bigg]\bigg)^{'}=0$$,\quad \quad\quad\quad   {EN}\quad &  \\ \\
$$\bigg(\rho^{n-1}h^{'}(1+\frac{h^{'2}}{4\beta^2 l^2})^{-1}\bigg)^{'}=0$$.\quad\quad\quad\quad \ {LN}\quad &
\end{array}
\right.
\end{equation}
Solving Eq. (\ref{equ3}) leads to the electromagnetic field as
\begin{eqnarray}\label{Fphir}
F_{\phi \rho}=h^{'}=\left\{
\begin{array}{ll}
$$l\beta\sqrt{-L_{W}(-\eta)}$$,\quad \quad\quad\quad \quad\quad\quad\quad \ {EN}\quad &  \\ \\
$$\frac{2ql^{n-2}}{\rho^{n-1}}(1+\sqrt{1-\eta})^{-1}$$,\quad\quad\quad\quad\quad  \ {LN}\quad &
\end{array}
\right.
\end{eqnarray}
where $\eta=\frac{q^2l^{2n-6}}{\beta^2 \rho^{2n-2}}$ and $q$ is the constant of integration. The Lambert function,  $L_{W}$, has the expansion as follows:
\begin{eqnarray}
L_{W}(x)=x-x^2+\frac{3}{2}x^3+...~ .
\end{eqnarray}
Expanding $F_{\phi\rho}$ for large $\beta$ leads to
\begin{eqnarray}
F_{\phi\rho}=\frac{ql^{n-2}}{\rho^{n-1}}+\left\{
\begin{array}{ll}
$$\frac{q^3 l^{3n-8}}{2\beta^2\rho^{3n-3}}+\mathcal{O}(\frac{1}{\beta^4})$$,\quad \quad\quad\quad\quad\quad\quad  \ {EN}\quad &  \\ \\
$$\frac{q^3 l^{3n-8}}{4\beta^2\rho^{3n-3}}+\mathcal{O}(\frac{1}{\beta^4})$$.\quad\quad\quad\quad\quad\quad\quad  \ {LN}\quad &
\end{array}
\right.
\end{eqnarray}
The first and the second terms are the magnetic brane electromagnetic fields in the presence of linear Maxwell theory in higher dimensions \cite{Taghi} and nonlinear electrodynamics respectively. As the four-vector potential $A_{\phi}$ depends just on the coordinate $\rho$, we would have $F_{\phi\rho}=-\partial_{\rho}A_{\phi}$. It can be obtained via solving $A_{\phi}=-\int{F_{\phi\rho}d\rho}$ as
\begin{eqnarray}\label{A1}
A_{\phi}=\left\{
\begin{array}{ll}
$$-\frac{n-1}{n-2}l\beta\big(\frac{l^{n-3}q}{\beta}\big)^{\frac{1}{n-1}}\big(-L_{W}(-\eta)\big)^{\frac{n-2}{2(n-1)}}\bigg\{ {}_2F_{1}\bigg(\big[\frac{n-2}{2(n-1)}\big]\,,\big[\frac{3n-4}{2(n-1)}\big]\,,-\frac{1}{2(n-1)}L_{W}(-\eta)\bigg)\\
-\frac{n-2}{n-1} \mathrm {exp}\Big[-\frac{1}{2(n-1)}L_{W}(-\eta)\Big]\bigg\}$$,\quad\quad\quad\quad\quad\quad\quad\quad\quad\quad\quad\quad\quad\quad\quad\quad\quad\quad\quad\quad\quad\quad\quad\quad\quad  \ {EN}\quad &  \\ \\
$$\frac{ql^{n-2}}{(n-2)\rho^{n-2}}{}_{3}F_{2}([\frac{n-2}{2(n-1)},\frac{1}{2},1]\,,[\frac{3n-4}{2(n-1)},2]\,,\eta)$$.\quad\quad\quad\quad\quad\quad\quad\quad\quad\quad\quad\quad\quad\quad\quad\quad\quad \quad \quad\quad  \ {LN}\quad &
\end{array}
\right.
\end{eqnarray}
As $\beta\rightarrow\infty$, for the four-vector potential of the Maxwell theory in $(n+1)$-dimensions  we have \cite{Taghi}
\begin{eqnarray}
A_{\phi}=\frac{ql^{n-2}}{(n-2)\rho^{n-2}}.
\end{eqnarray}
Eqs. (\ref{equ2}) and (\ref{Fphir}) yield to the  quintic equation for $\Psi$ as
\begin{eqnarray}\label{quintic1}
\hat{\mu}_5 \Psi^5+\hat{\mu}_4 \Psi^4+\hat{\mu}_3 \Psi^3+\hat{\mu}_2 \Psi^2+\Psi+\kappa=0.
\end{eqnarray}
Here $\kappa$ is
\begin{eqnarray}\label{kappa1}
\kappa&=&-\frac{2\Lambda l^2}{n(n-1)}-\frac{M}{(n-1)\rho^n}+\nonumber\\
&&\left\{
\begin{array}{ll}
$$-\frac{4 l^2\beta^2}{n(n-1)}+\frac{4(n-1)\beta q l^{2}}{n(n-2)\rho^n}(\frac{q}{\beta})^{\frac{1}{n-1}}\big(L_{W}(\eta)\big)^{\frac{n-2}{2(n-1)}}\times{}_2 F_{1}\Big([\frac{n-2}{2(n-1)}]\,,[\frac{3n-4}{2(n-1)}]\,,-\frac{1}{2(n-1)}L_{W}(\eta)\Big)\\-\frac{4\beta q l^{2}}{(n-1)\rho^{n-1}}[L_{W}(\eta)]^{\frac{1}{2}}\times\Big[1-\frac{1}{n}\Big(L_{W}(\eta)\Big)^{-1}\Big]$$,\quad \quad\quad\quad\quad\quad \quad\quad\quad\quad\quad\quad\quad  \ {EN}\quad &  \\ \\
$$\frac{8(2n-1)}{n^2(n-1)}\beta^2 l^2[1-\sqrt{1+\eta}]+\frac{8(n-1)q^2l^{2}}{n^2 (n-2)\rho^{2n-2}}{}_2 F_{1}\Big([\frac{n-2}{2(n-1)},\frac{1}{2}]\,,[\frac{3n-4}{2(n-1)}]\,,-\eta\Big)\\
-\frac{8}{n(n-1)}l^2\beta^2 \mathrm {ln}[\frac{2\sqrt{1+\eta}-2}{\eta}]$$,\quad\quad\quad\quad\quad\quad\quad\quad\quad\quad\quad\quad\quad\quad\quad\quad\quad\quad \quad\quad\quad  \ {LN}\quad &
\end{array}
\right.
\end{eqnarray}
and $M$ is the constant of integration which is related to the mass of the
spacetime.
\begin{figure}
    \centering
    \subfigure[LN]{\includegraphics[scale=0.8]{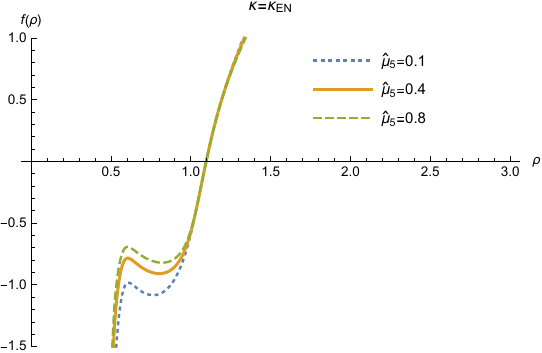}\label{figa}}\hspace*{.2cm}
    \subfigure[EN]{\includegraphics[scale=0.8]{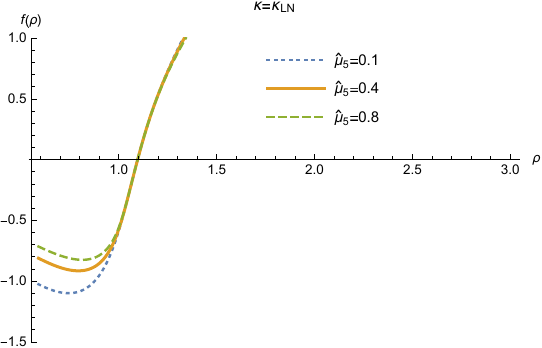}\label{figb}}\caption{ $f(\rho)$ versus $\rho$ for different values of  $\hat{\mu}_{5}$ (0.1, 0.4 and 0.8) with $k=0$, $n=4$, l=1, $\hat{\mu}_{5}=0.1$, $\hat{\mu}_{4}=-0.01$, $\hat{\mu}_{3}=0.4$, $\hat{\mu}_{2}=-0.01$, $M=1$, $q=2$, $\beta=10$.}\label{figure1}
\end{figure}
\begin{figure}
    \centering
    \subfigure[LN]{\includegraphics[scale=0.8]{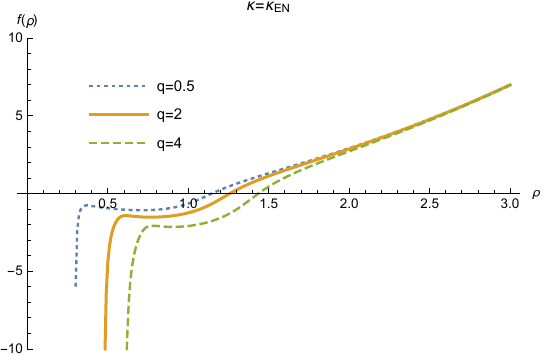}\label{figc}}\hspace*{.2cm}
    \subfigure[EN]{\includegraphics[scale=0.8]{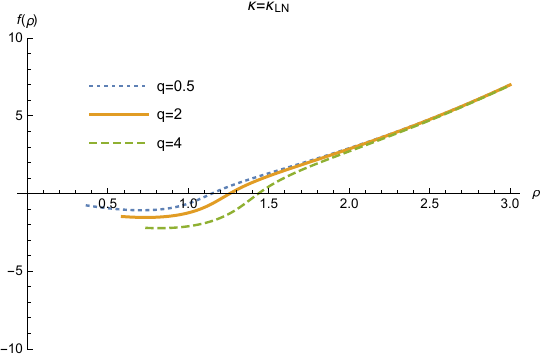}\label{figd}}\caption{ $f(\rho)$ versus $\rho$ for different values of $q$ (0.5, 2 and 4) with $k=0$, $n=4$, $\hat{\mu}_{5}=0.01$, $\hat{\mu}_{4}=-0.01$, $\hat{\mu}_{3}=0.4$, $\hat{\mu}_{2}=-0.01$, $M=1$, $\beta=10$.}\label{figure2}
\end{figure}

As a quintic equation has no analytical solution we need to go for a numerical one. Here we plot the behavior of $f(\rho)$ versus $\rho$ for different values of  charge and quintic quasi topological gravity parameter.

According to Figs. (\ref{figure1}) and (\ref{figure2}), we can see that $f(\rho)$ has one real positive root (here $r_+ = ...$) where the metric function is negative for $\rho<r_+$ and positive for $\rho>r_+$. Therefore the signature of the metric changes from $(-,+,+,+,...)$ to $(-,-,-,+,...)$ in $0<\rho<r_+$ interval. In line with this interpretation, it is clear that it is not possible to extend the spacetime to $\rho<r_+$. To avoid this improper extension, we have to introduce a suitable radial variable $r$ as $r^2=\rho^2-r_+^2$, in order for the space to be started at $r=0$. Therefore the metric is
\begin{eqnarray}\label{metric2}
ds^2=-\frac{r^2+r_{+}^2}{l^2}dt^2+\frac{r^2 dr^2}{(r^2+r_{+}^2)f(r)}+l^2 g(r)d \phi^2+\frac{r^2+r_{+}^2}{l^2}dX^2.
\end{eqnarray}
By this transformation, $r$ has the range $0\leq r<\infty$ which $f(r)$ is positive and real for $0< r<\infty$, and zero for $r=0$.
This transformation also makes $\kappa$ to be changed as:

\begin{eqnarray}\label{kappa2}
\kappa&=& 1-\frac{M}{(n-1)(r^2+r_{+}^2)^{\frac{n}{2}}}\nonumber\\
&+&\left\{
\begin{array}{ll}
$$-\frac{4 l^2\beta^2}{n(n-1)}-\frac{4(n-1)\beta q l^{n-1}}{n(n-2)(r^2+r_{+}^2)^{\frac{n}{2}}}\Big(\frac{l^{n-3}q}{\beta}\Big)^{\frac{1}{n-1}}\Big(-L_{W}(-\eta)\Big)^{\frac{n-2}{2(n-1)}}\times{}_2 F_{1}\Big([\frac{n-2}{2(n-1)}]\,,[\frac{3n-4}{2(n-1)}]\,,\\
-\frac{1}{2(n-1)}L_{W}(-\eta)\Big)+\frac{4\beta q l^{n-1}}{(n-1)(r^2+r_{+}^2)^{\frac{n-1}{2}}}[-L_{W}(-\eta)]^{\frac{1}{2}}\times\Big[1+\frac{1}{n}\Big(-L_{W}(-\eta)\Big)^{-1}\Big]
$$,\quad  \ {EN}\quad &  \\ \\
$$\frac{8(2n-1)}{n^2(n-1)}\beta^2 l^2[1-\sqrt{1-\eta}]-\frac{8(n-1)q^2l^{2n-4}}{n^2 (n-2)(r^2+r_{+}^2)^{n-1}}{}_2 F_{1}\Big([\frac{n-2}{2(n-1)},\frac{1}{2}]\,,[\frac{3n-4}{2(n-1)}]\,,\eta\Big)\\-\frac{8}{n(n-1)}l^2\beta^2 \mathrm {ln}[\frac{2-2\sqrt{1-\eta}}{\eta}]$$, \quad\quad\quad\quad \quad\quad\quad\quad \quad\quad\quad\quad \quad\quad\quad\quad \quad\quad\quad\quad \quad\quad\quad  \ {LN}\quad &
\end{array}
\right.
\end{eqnarray}
where $\eta=\frac{q^2l^{2n-6}}{\beta^2(r^2+r_{+}^2)^{n-1}}$.
Calculation gives the Kretschmann scalar as
\begin{eqnarray}
{\cal K} = R_{\mu\nu\alpha\beta}R^{\mu\nu\alpha\beta} = f^{\prime\prime 2}+~\cdots,
\end{eqnarray}
where $f^{\prime\prime}$ is the second derivative of the metric.
From the Kretschmann scalar one can realize that at $\rho=0$ there is a singularity, but as we discussed, the $\rho=0$ region cannot be included in the admissible space. Therefore the magnetic brane solution is free of singularity. As there is no analytical solution for the quintic equation (\ref{quintic1}) , we plot it numerically.

\begin{figure}
  \centering
  \subfigure[LN]{\includegraphics[scale=0.8]{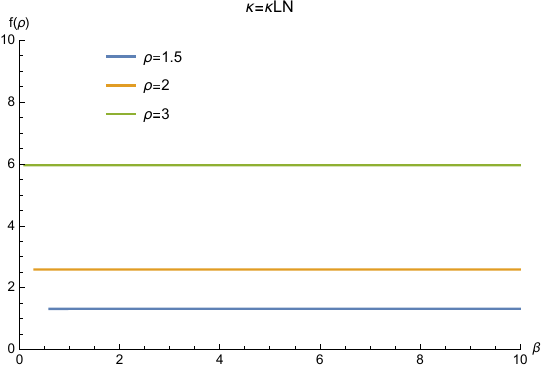}\label{fig4a}}\hspace*{.2cm}
  \subfigure[EN]{\includegraphics[scale=0.8]{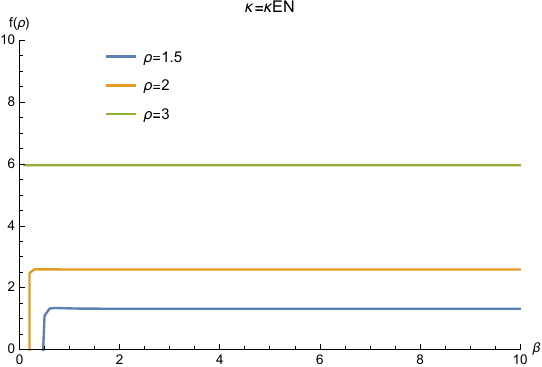}\label{fig4b}}\caption{ $f(\rho)$ versus $\beta$ for different values of $\rho$ (1.5, 2 and 3) with $k=0$, $n=4$, $\hat{\mu}_{5}=0.01$, $\hat{\mu}_{4}=-0.01$, $\hat{\mu}_{3}=0.4$, $\hat{\mu}_{2}=-0.01$ and $M=1$.}\label{figure3}
\end{figure}
In Figs (\ref{figure1}) and (\ref{figure2}) we plot the metric
function versus  $\rho$  and in Fig (\ref{figure3}) versus $\beta$ for different values of $\hat{\mu_5}$,
$q$ and $\rho$ respectively for two
different cases of $EN$ and $LN$.

Due to the previous explanation about the metric function, we can see from Fig. (\ref{figure1}) that there is an $r_{+}$ where for $\rho<r_{+}$, $f(\rho)$ is negative which is unacceptable. Form Fig (\ref{figure2}) it is obvious that keeping other parameters constant, $r_{+}$ increases with $q$. It's noteworthy to mention that for a specific $q$, $r_+$ does not depend on nonlinear parameter, i.e. there is no difference between either of $LN$ or $EN$ nonlinear electrodynamics that can be seen from Fig. (\ref{figure3}).
According to Fig. (\ref{figure3}) one can find out that for a specific $\rho$, the function $f(\rho)$ is a constant function of $\beta$ which is imaginary for $\beta<\beta_{min}$ and is real for $\beta$ larger than $\beta_{min}$. The value of $\beta_{min}$, where the plot ends at, decreases as $\rho$ increases.
Also Fig. (\ref{figure1}) tells us that the function $f(\rho)$ is not sensitive to $\hat{\mu_5}$  for $\rho > r_+$ and has a similar behavior with variation of this parameter.

\begin{figure}
	\centering
	\subfigure[LN]{\includegraphics[scale=0.6]{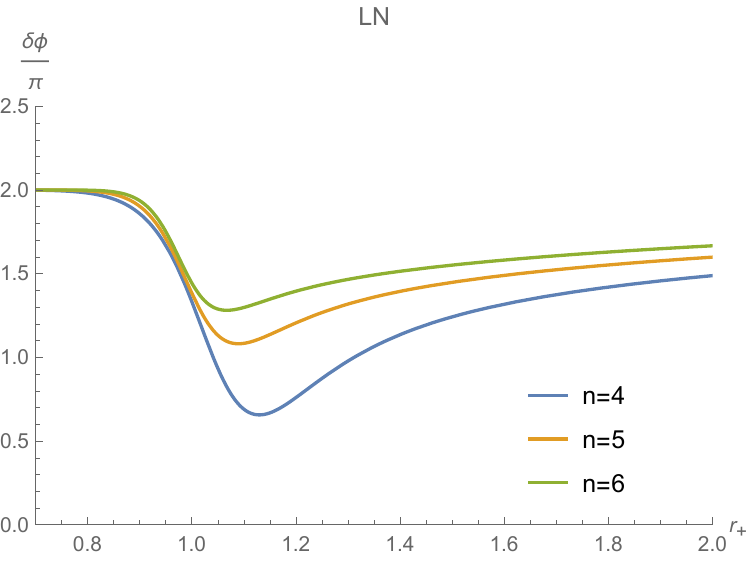}\label{phivsrp4nLN}}\hspace*{.2cm}
	\subfigure[EN]{\includegraphics[scale=0.6]{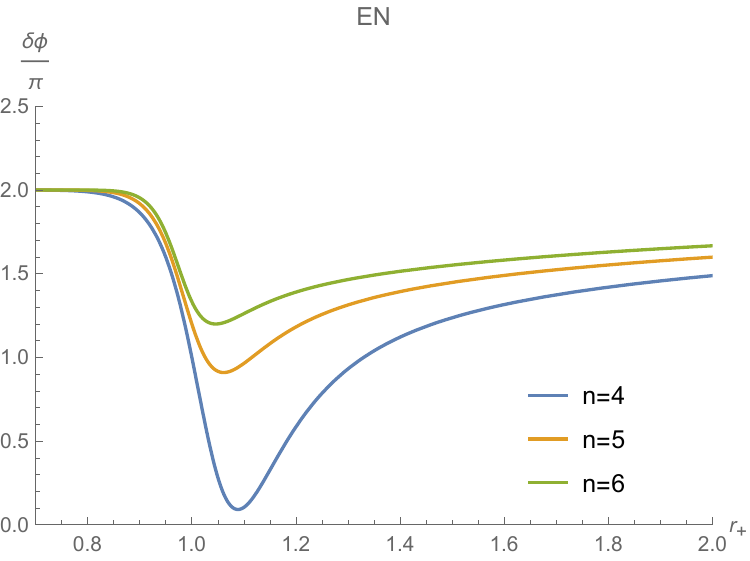}\label{phivsrp4nEN}}\caption{ $\delta\phi$ versus $r_{+}$ for different values of $n$  with $k=0$, $l=1$, $q=3$ and $\beta = 5$. The value of $r_{+_{min}}$ decreases as $n$ increases.}\label{phivsrp4n}
\end{figure}

\begin{figure}
	\centering
	\subfigure[LN]{\includegraphics[scale=0.6]{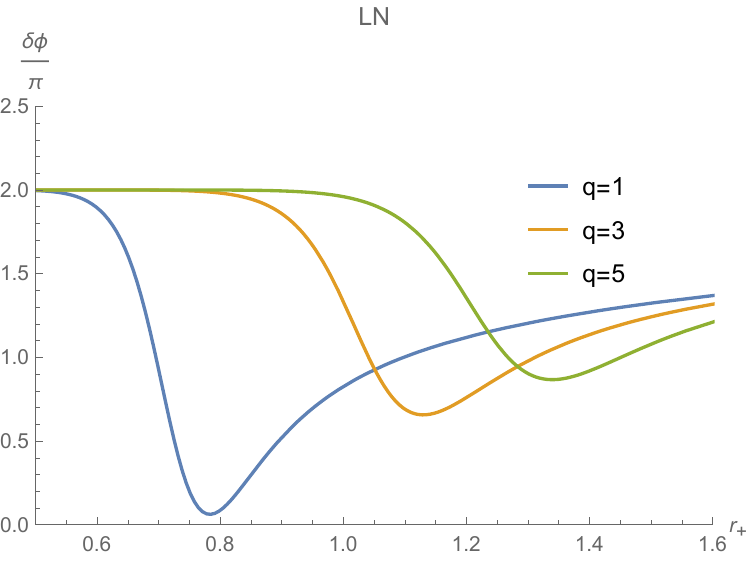}\label{phivsrp4qLN}}\hspace*{.2cm}
	\subfigure[EN]{\includegraphics[scale=0.6]{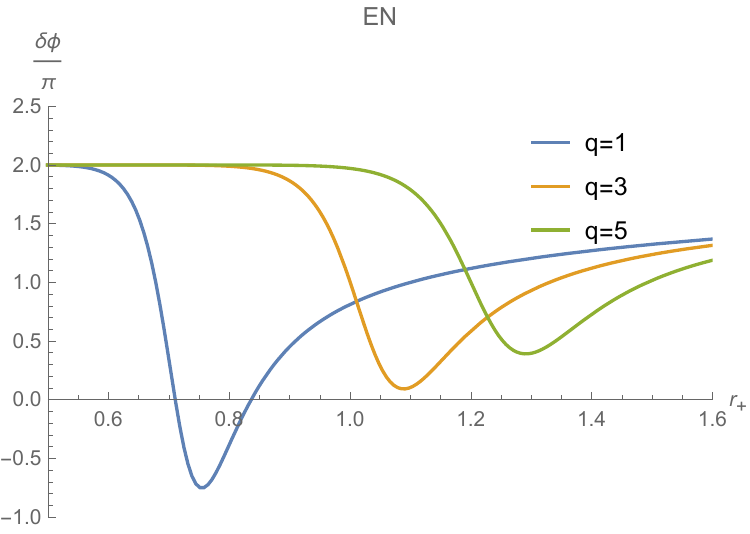}\label{phivsrp4qEN}}\caption{ $\delta\phi$ versus $r_{+}$ for different values of $q$  with $k=0$, $l=1$, $n=4$ and $\beta = 5$. The value of $r_{+_{min}}$ increases as $q$ increases.}\label{phivsrp4q}
\end{figure}

\begin{figure}
	\centering
	\subfigure[LN]{\includegraphics[scale=0.6]{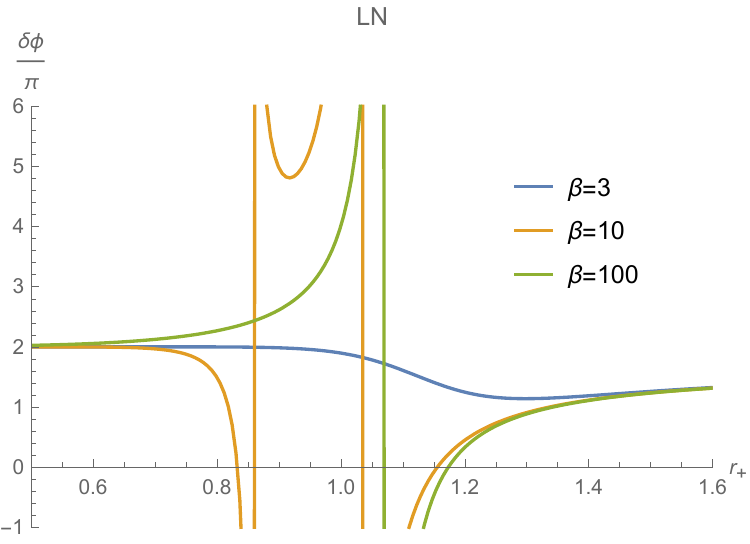}\label{phivsrp4betaLN}}\hspace*{.2cm}
	\subfigure[EN]{\includegraphics[scale=0.6]{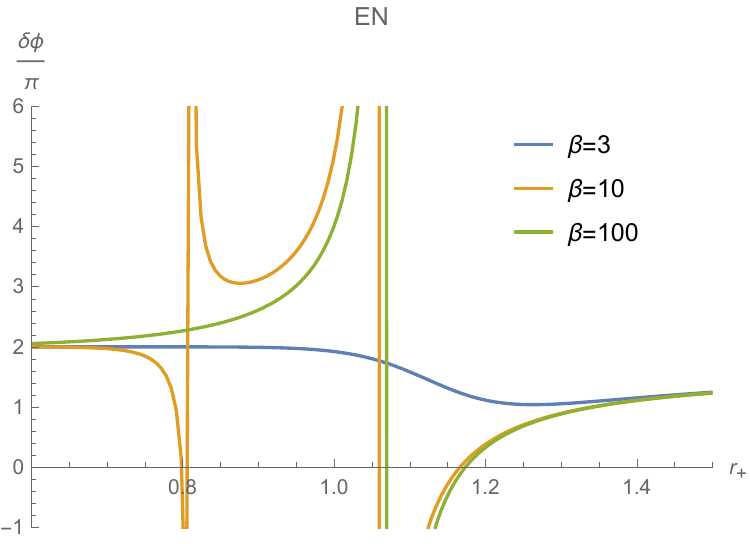}\label{phivsrp4betaEN}}\caption{ $\delta\phi$ versus $r_{+}$ for different values of $\beta$  with $k=0$, $l=1$, $q=3$ and $n = 4$. }\label{phivsrp4beta}
\end{figure}

Although the Kretschmann scalar does not diverge in $0\leq r<\infty$, but it can be shown that there is a conical singularity at $r=0$.
The circumference/radius ratio can be used to study the conic geometry as well.
Taylor expanding around $r=0$,
\begin{eqnarray}
f(r)=f(r)\mid_{r=0}+r\frac{df(r)}{dr}\mid_{r=0}+\frac{r^2}{2}\frac{d^2 f(r)}{dr^2}\mid_{r=0}+\mathcal{O}(r^3).
\end{eqnarray}
one can show that
\begin{eqnarray}\label{phi1}
\lim_{r \to 0} \big( \sqrt{\frac{g_{\phi\phi}}{g_{rr}} } \big)^{-1}= .... \neq 1.
\end{eqnarray}
Therefore, as $r$ approaches to zero, it can be claimed that the limit of circumference/radius ratio is not $2\pi$, which yields to the conclusion that the spacetime has a conical singularity at $r=0$. It can be removed if the coordinate $\phi$ has the following period:
\begin{eqnarray}
Period_{\phi}=2\pi \bigg(\mathrm{lim}_{r\rightarrow 0}\bigg(\frac{1}{r}\sqrt{\frac{g_{\phi\phi}}{g_{rr}}}\bigg)\bigg)^{-1}=2\pi(1-4\tau),
\end{eqnarray}
where $\tau$ is obtained as
\begin{eqnarray}\label{tau}
\tau=\frac{1}{4}\bigg[1-\frac{2l}{r_{+}^3}\bigg(\frac{d^2 f(r)}{dr^2}\Bigg|_{r=0}\bigg)^{-1}\bigg].
\end{eqnarray}
In other words, the near origin limit of the metric describes a local flat spacetime which has a conical singularity at $r=0$ with a deficit angle $\delta\phi=8\pi\tau$.
From Eq. (\ref{tau}) we can see that the value of deficit angle $\delta\phi$ depends on the parameters $n$, $q$ and $\beta$. To clarify this point we plot $\delta\phi$ as a function of $r_{+}$ for different values of the above parameters in Figs. (\ref{phivsrp4n}), (\ref{phivsrp4q}) and (\ref{phivsrp4beta}).
The effect of dimension is shown in Fig. (\ref{phivsrp4n}). It is evident that for both EN and LN cases the minimum of curve, $r_{+_{min}}$, is a decreasing function of dimension but the value of corresponding deficit angle increases. Also for different values of $q$, one can find out from Fig. (\ref{phivsrp4q}) that both $r_{+_{min}}$ and its corresponding deficit angle increases with $q$.

\subsection*{\textit{Divergency}}

The nonlinearity decreases with the increase of parameter $\beta$. So the large values of $\beta$ indicates small nonlinear behavior which can be treated as a perturbation. According to Fig. (\ref{phivsrp4beta}), for small values of $\beta$ (high nonlinearity, $\beta=3$) there is no divergency at all. But for large values of $\beta$ ($\beta = 10$) one can see three different behaviors considering different intervals of $r_{+}$. Theses regions can be identified with $r_{Div1}$ and $r_{Div2}$ in the plot. For $0<r_{+}<r_{Div1}$ the deficit angle is a decreasing function of $r_{+}$ and there is a divergency at $r_{+}=r_{Div1}$. For $r_{Div1}<r_{+}<r_{Div2}$, the deficit angle is a real positive function of $r_{+}$ with a minimum and we see the second divergency at $r_{+}=r_{Div2}$. For $r_{+}>r_{Div2}$, the deficit angle is an increasing function which has a zero at $r_{+}=r_0$. For very high values of $\beta$ ($\beta = 100$), which corresponds to linear Maxwell theory, there is just the second divergency and the plot is independent of the values of $\beta$.

\subsection*{\textit{Energy Conditions}}
\begin{figure}
	\centering
	\subfigure[LN]{\includegraphics[scale=0.6]{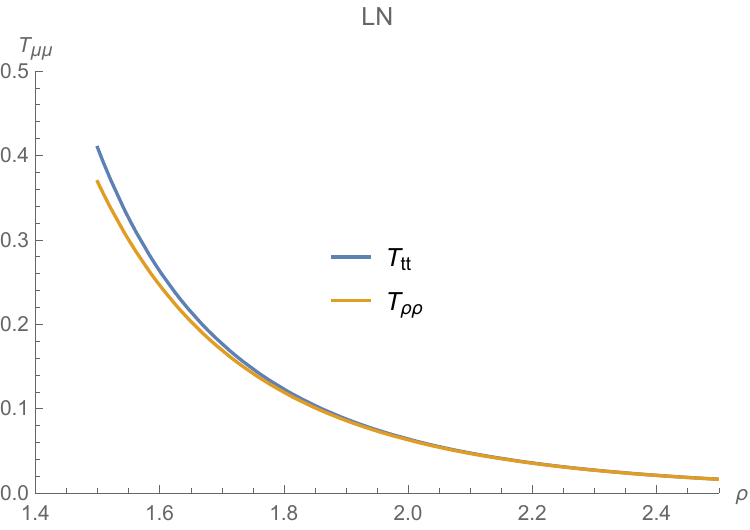}\label{ELN}}\hspace*{.2cm}
	\subfigure[EN]{\includegraphics[scale=0.6]{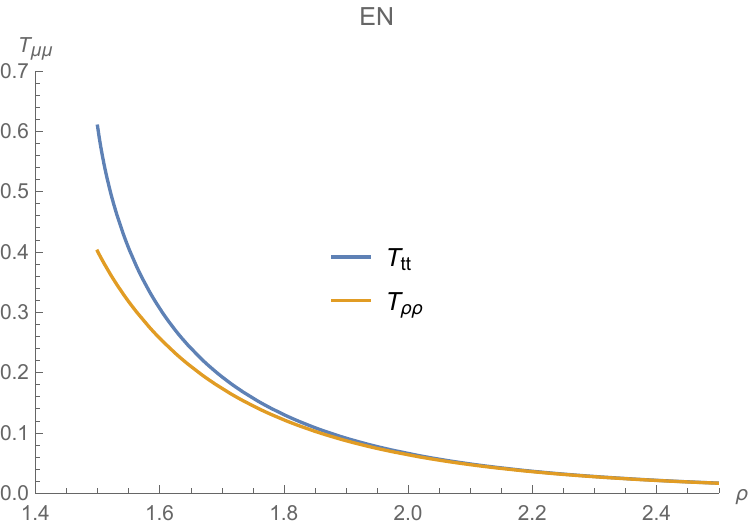}\label{EEN}}\caption{Plot of energy-momentum tensor elements, $T_{_{\widehat{t}\widehat{t}}}$ and $T_{_{\widehat{\rho}\widehat{\rho}}}$, vs $\rho$ with $k=0$, $n=4$ and $\beta=1$. For both LN and EN cases it is obvious that $T_{_{\widehat{t}\widehat{t}}} \ge 0$ and also $T_{_{\widehat{t}\widehat{t}}} \ge T_{_{\widehat{\rho}\widehat{\rho}}}$, which represent the dominant energy condition.}\label{EngCond}
\end{figure}

Now we are going to discuss the energy conditions for magnetic brane solutions. We first diagonalize the energy-momentum tensor and to simplify the mathematics we use the following orthonormal contravariant basis:
\begin{equation}
\mathbf{e}_{\widehat{t}}=\frac{l}{\rho}\frac{\partial }{\partial t},\;~ \;~
\mathbf{e}_{\widehat{\rho}}=f^{1/2}\frac{\partial }{\partial \rho},\;~\;~ \mathbf{e%
}_{\widehat{\phi }}=\frac{1}{ lf^{1/2}}\frac{\partial }{\partial
	\phi },\;~\;~\mathbf{e}_{\widehat{x_i }}=\rho^{-1}\frac{\partial }{\partial
	x_i }~.  \label{base}
\end{equation}
The diagonal elements of the energy-momentum tensor are obtained as:
\begin{eqnarray}
T_{_{\widehat{t}\widehat{t}}} =-T_{_{\widehat{x}\widehat{x }
}}&=&-T_{_{\widehat{y}\widehat{y}
}}=\left\{
\begin{array}{ll}
2 \beta ^2 \left( 1 - e^{\frac{L_W(-\eta )}{2}}\right), & \;~{EN} \\
4 \beta ^2 \ln \left(\frac{ 2}{ \sqrt{1-\eta }+1}\right), & \;~{LN}%
\end{array}
\right. ,  \label{Tab1} \\
\;T_{_{\widehat{\rho}\widehat{\rho}}} &=&T_{_{\widehat{\phi }\widehat{\phi }%
}}=\left\{
\begin{array}{ll}
2 \beta ^2 \left(e^{\frac{L_W(-\eta )}{2}} [1-L_W(-\eta )]-1\right), &
\;~{EN} \\
4 \beta ^2 \Bigg[\frac{\eta }{\sqrt{1-\eta }+1}-\ln \left(\frac{2}{\sqrt{1-\eta }+1}\right)\Bigg], & \;~{LN}%
\end{array}
\right. .  \label{Tab2}
\end{eqnarray}
For both EN and LN cases and all ranges of $\rho$, one can verify that the dominant energy conditions
\begin{eqnarray}
T_{_{\widehat{t}\widehat{t}}} &\ge& 0, \nonumber\\
T_{_{\widehat{t}\widehat{t}}} &\ge& |T_{_{\widehat{\alpha}\widehat{\alpha}}}|~,
\end{eqnarray}
are satisfied, in which $\alpha$ represents all spacelike coordinates $\rho$, $\phi$, $x$ and $y$. (See Fig. (\ref{EngCond}) for more clarification.)

\begin{figure}
	\centering
	\subfigure[LN]{\includegraphics[scale=0.6]{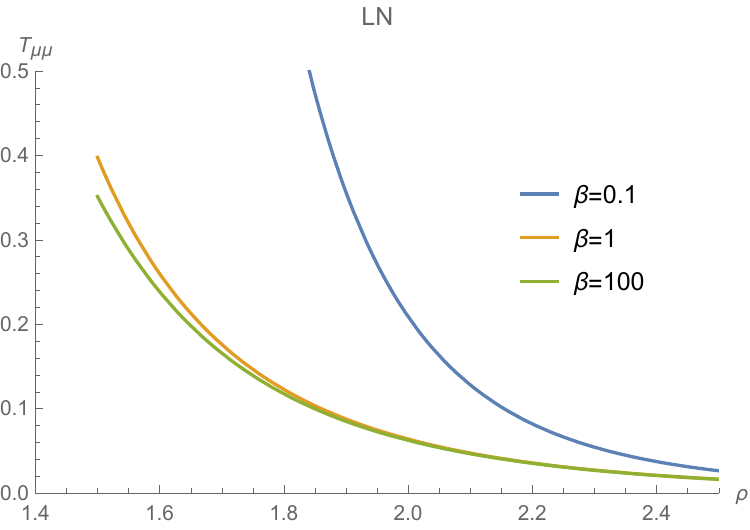}\label{EnLN}}\hspace*{.2cm}
	\subfigure[EN]{\includegraphics[scale=0.6]{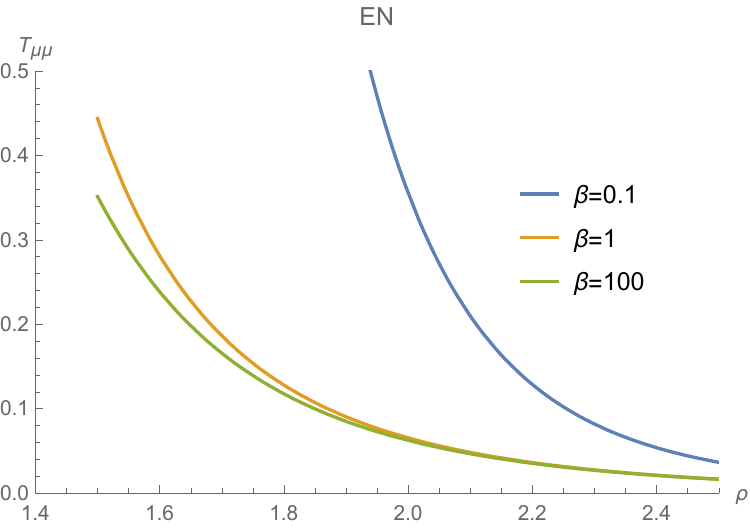}\label{EnEN}}\caption{Energy density curve for different values of nonlinear parameter $\beta$. For both LN and EN cases, one can see that the difference between curves decreases as the value of $\beta$ increases which is analogous to less nonlinear contribution. }\label{Energy}
\end{figure}

To investigate the effect of nonlinearity on energy density of spacetime, we expand $T_{_{\widehat{t}\widehat{t}}}$ near the linear region (large $\beta$) which yields
\begin{equation}
T_{_{\widehat{t}\widehat{t}}}= \frac{q^2 l^{2 n-6} }{\rho^{2 n-2}} +\left\{
\begin{array}{ll}
\dfrac{3 q^4 l^{4 (n-3)} }{8 \beta ^2 \rho^{4n-4}} + {\cal O}(\frac{1}{\beta^4}), & \;~LN \\
\dfrac{3 q^4 l^{4 (n-3)}}{4 \beta ^2  \rho^{4n-4}}+ {\cal O}(\frac{1}{\beta^4}), & \;~EN%
\end{array}
\right. .  \label{Tbast}
\end{equation}
The first term is the linear Maxwell contribution which for the usual 4D spacetime ($n=3$) reduces to the ordinary $q^2/\rho^4$ behavior. The second terms are the LN and EN corrections and one can easily see that the contribution of EN case is twice as that of LN.
As can be seen from Fig. (\ref{Energy}), for $\beta \ge 1$ the effect of nonlinearity reduces considerably with respect to the case that $\beta < 1$.

\section{A Class of Spinning Solutions}\label{spin}
In this section we generalize the static spacetime to rotating solutions. As we know, the rotational group in $(n+1)$ dimensions is $SO(n-2)$ with $\lfloor \frac{(n-2)}{2} \rfloor$ independent parameters, where $\lfloor x \rfloor$ gives the greatest integer less than or equal to $x$. The magnetic rotational solution with $k\leq \lfloor \frac{(n-2)}{2} \rfloor $ is

\begin{eqnarray}
ds^{2} &=&-\frac{r^{2}+r_{+}^{2}}{l^{2}}\left( \Xi dt-{{\sum_{i=1}^{k}}}%
a_{i}d\phi ^{i}\right) ^{2}+f(r)\left( \sqrt{\Xi ^{2}-1}dt-\frac{\Xi }{\sqrt{%
        \Xi ^{2}-1}}{{\sum_{i=1}^{k}}}a_{i}d\phi ^{i}\right) ^{2}  \notag \\
&&+\frac{r^{2}dr^{2}}{(r^{2}+r_{+}^{2})f(r)}+\frac{r^{2}+r_{+}^{2}}{%
    l^{2}(\Xi ^{2}-1)}{\sum_{i<j}^{k}}(a_{i}d\phi _{j}-a_{j}d\phi _{i})^{2}+%
\frac{r^{2}+r_{+}^{2}}{l^{2}}dX^{2},  \label{Metric3}
\end{eqnarray}%
where $\Xi = \sqrt{1+\sum_{i=1}^{k}\frac{a_i^2}{l^2}}$, $dX^2$ is the Euclidean metric on the $(n-k-3)$ dimensions submanifold with volume $V_{n-k-3}$ and $f(r)$ is the function introduced in non-rotational metric (\ref{metr}).
We must note that the nonzero components of the electromagnetic field are

\begin{equation}
F_{rt}=-\frac{(\Xi ^{2}-1)}{\Xi a_{i}} F_{r\phi _{i}}.
\end{equation}
Again, we should note that although this rotating spacetime has no curvature
singularity and horizon, it has a conical singularity at $r=0$.
Although it is necessary to mention that considering $a_i=0$ or $\Xi  = 1$, one leads to the static solutions in the previous section.

\section{Physical Properties of the Magnetic Solutions}\label{struc}

The action $I_G + I_b$ would diverge on solutions. To remove the divergence one can add a counterterm, $I_{\text{ct}}$, that is a function of the invariants of the boundary curvature as
\begin{eqnarray}
I_{ct}=-\frac{1}{8\pi}\int_{\partial\mathcal{M}} d^{n}x \sqrt{-\gamma}\frac{(n-1)}{l_{eff}},
\end{eqnarray}
where $l_{eff}$ is a scale length factor which is a function of $l$ and gravitational coupling constants, and has to be reduced to $l$ in the absence of higher curvature terms.
$l_{eff}$
\  is chosen as
\begin{equation}
l_{eff}=\frac{630\Psi _{\infty }^{1/2}}{(35\hat{\mu _{5}}\Psi _{\infty
    }^{5}-45\hat{\mu _{4}}\Psi _{\infty
    }^{4}-63\hat{\mu _{3}}\Psi _{\infty }^{3}-105\hat{\mu _{2}}\Psi _{\infty
    }^{2}+315\Psi _{\infty }+315)}l,  \label{leff}
\end{equation}%
where $\Psi _{\infty }$ is the limit of $\Psi $ at infinity for our solution.
Considering a spacelike
surface $\mathcal{B}$ in $\partial \mathcal{M}$ with metric $\sigma_{ij}$, one can write the boundary metric as 
\begin{eqnarray}
	\gamma _{ab}dx^{a}dx^{b}=-N^2 dt^2+\sigma_{ij}(d\phi^{i}+V^{i} dt)(d\phi^{j}+V^{j} dt),
\end{eqnarray}
in ADM form. The coordinates $\phi^{i}$,  $V^{i}$ and $N$ are the angular variables, shift function and the lapse function respectively.

Employing the new finite action, the quasilocal conserved quantity, which is the total charge per unit volume $V_{n-1}$ is
\begin{eqnarray}
\mathcal{Q}(\xi)=\int_{\mathcal{B}}d^{n-1} \phi \sqrt{\sigma} T_{ab} n ^{a} \xi^{b}.
\end{eqnarray}
It can be obtained by calculating the finite stress tensor $T_{ab}$. Here $n^{a}$, $\sigma$ and  $\xi^{b}$ are the unit vector normal to
the boundary $\mathcal{B}$, the determinant of the metric $\sigma_{ij}$ and the Killing vector field on the boundary respectively. Considering the boundaries with timelike $(\xi=\partial/\partial t)$ and rotational
($\varsigma=\partial/\partial  \phi$) Killing vector fields, as the boundary $\mathcal{B}$ goes to infinity, the mass and the angular momentum per unit volume $V_{n-1}$ of this black brane are obtained as
\begin{eqnarray}\label{mass}
M=\int_{\mathcal{B}} d^{n-1}\phi \sqrt{\sigma}T_{ab}n^{a}\xi^{b}=\frac{(n\Xi^2-1)}{16\pi(n-1)l^{n-1}}m,
\end{eqnarray}
\begin{eqnarray}\label{angular}
J_{i}=\int_{\mathcal{B}} d^{n-1}\phi \sqrt{\sigma}T_{ab}n^{a}\varsigma_{i}^{b}=\frac{1}{16\pi(n-1)l^{n-1}}n\Xi m a_{i}.
\end{eqnarray}
It is clear that for $a_{i}=0$ (or $\Xi=1$), the angular momentum $J_{i}$ vanishes and we can calculate the
total mass per unit volume $V_{n-1}$ as
\begin{eqnarray}
M_{total} =\frac{M}{4(n-1)},
\end{eqnarray}
which can be obtained by using the relevant dedicated Killing vector, $\xi=\partial /\partial t$.
As we imposed the limit in which the boundary $\mathcal{B}$ is infinite, the obtained mass would be finite.

Now we proceed to evaluate this magnetic brane electric charge.
Considering the projection of the electromagnetic field tensors on the special hypersurfaces with normal
$u^0 =\frac{1}{N}$,
 $u^r=0$
 and
 $u^i=-\frac{N^i}{N} $,
 one can calculate the electric charge of the spacetime with a longitudinal magnetic field. Therefore the electric field is obtained as follows:
\begin{eqnarray}
E^{u}=g^{\mu\rho}F_{\rho\nu}u^{\nu}.
\end{eqnarray}
The electric charge per unit volume $V_{n-1}$ is zero as indicated from evaluation of the electromagnetic field at infinity.
This, in turn, causes the electric field to be zero.
One can evaluate the electric field as the magnetic brane rotates at least once, and as we consider just static magnetic branes, therefore, its electric field and consequently its relevant electric charge vanish accordingly.\\

\section{Summary and Concluding Remarks}\label{result}

In this work we introduced a surface term which makes the quintic quasitopological gravity well defined and also obtained the counterterm which causes the action to be finite.
At first we considered an appropriate static metric to find the horizonless magnetic solutions. We numerically analyzed the static magnetic layer solutions in $(n+1)$ dimensions in the presence of quasitopological gravity up to quintic order with two fields of nonlinear electrodynamics, the exponential and logarithmic ones. The characteristic of these solutions is that they don't have any horizon or singularity. Rather they got a conical singularity at $r=0$ with the deficit angle $\delta\phi$.

We reached to this conclusion that the parameters of the quasitopological gravity have no effect on this deficit angle. It comes from the fact that the second derivatives if the metric function does not depend on the quasitopological coefficients that is the consequence of the geometrical properties for the  hypersurface $t=\text{const}$ and $r=\text{const}$.

We also investigated the effect of non-linear electromagnetic fields. The two $EN$ and $LN$ models, being two versions of the $BI$ model, have different natures. We saw that there is a minimum value for the nonlinearity parameter, ($\beta_{min}$) which for $\beta\leq\beta_{min}$, the function $f(\rho)$ is not real valued. This is because of the Lambert function behavior which appears in $EN$ and $LN$ electromagnetic field.

We found out for the rotating magnetic layer solutions, in addition to a magnetic field, there would also be an electric field. Using the counterterm method, we calculated the electric charge, mass and angular momentum of the rotating magnetic layer. We realized that the electric charge is proportional to the rotating factor and in the static case where $\Xi=1$, it would be zero.

\section*{Acknowledgment}
AB and ARO would like to thank Jahrom University.  ARO is grateful to Institute for Research in Fundamental Sciences (IPM) for their warm hospitality,  where a part of this work was completed.

\section{APPENDIX}\label{app} 

In Eq. (\ref{quartic}), the coefficients $b_{i}$ are:
\begin{table}[!hpt]
\begin{center}
\begin{tabular}{|c|c|c|}
\hline \textbf{Label}&\textbf{Term}&$b_i$\\ \hline \hline
$\mathcal{L}^{(4)}_{1}$ &$R_{abcd}R^{cdef}R^{hg}{{}_{ef}}R_{hg}{{}^{ab}}$ &$-(n-1)(n^7-3n^6-29n^5+170n^4-349n^3+348n^2-180n+36)$ \\
  \hline
 $\mathcal{L}^{(4)}_{2}$ &$R_{abcd}R^{abcd}R_{ef}{{R}^{ef}}$& $-4(n-3)(2n^6-20n^5+65n^4-81n^3+13n^2+45n-18)$ \\
  \hline
  $\mathcal{L}^{(4)}_{3}$  &$RR_{ab}R^{ac}R_c{{}^b}$& $-64(n-1)(3n^2-8n+3)(n^2-3n+3)$ \\
 \hline
  $\mathcal{L}^{(4)}_{4}$ &$(R_{abcd}R^{abcd})^2$ & $-(n^8-6n^7+12n^6-22n^5+114n^4-345n^3+468n^2-270n+54) $\\
 \hline
 $\mathcal{L}^{(4)}_{5}$ &$R_{ab}R^{ac}R_{cd}R^{db}$ &$16(n-1)(10n^4-51n^3+93n^2-72n+18)$\\
  \hline
 $\mathcal{L}^{(4)}_{6}$ &$RR_{abcd}R^{ac}R^{db}$ &$-32(n-1)^2(n-3)^2(3n^2-8n+3)$\\
  \hline
 $\mathcal{L}^{(4)}_{7}$ &$R_{abcd}R^{ac}R^{be}R^d{{}_e}$ &$64(n-2)(n-1)^2(4n^3-18n^2+27n-9)$\\
  \hline
 $\mathcal{L}^{(4)}_{8}$ &$R_{abcd}R^{acef}R^b{{}_e}R^d{{}_f}$ &$-96(n-1)(n-2)(2n^4-7n^3+4n^2+6n-3)$\\
  \hline
 $\mathcal{L}^{(4)}_{9}$ &$R_{abcd}R^{ac}R_{ef}R^{bedf}$ &$16(n-1)^3(2n^4-26n^3+93n^2-117n+36)$\\
  \hline
 $\mathcal{L}^{(4)}_{10}$ &$R^4$ &$n^5-31n^4+168n^3-360n^2+330n-90$\\
  \hline
 $\mathcal{L}^{(4)}_{11}$ &$R^2 R_{abcd}R^{abcd}$ &$2(6n^6-67n^5+311n^4-742n^3+936n^2-576n+126)$\\
  \hline
 $\mathcal{L}^{(4)}_{12}$ &$R^2 R_{ab}R^{ab}$ &$8(7n^5-47n^4+121n^3-141n^2+63n-9)$\\
  \hline
 $\mathcal{L}^{(4)}_{13}$ &$R_{abcd}R^{abef}R_{ef}{{}^c{{}_g}}R^{dg}$ &$16n(n-1)(n-2)(n-3)(3n^2-8n+3)$\\
  \hline
 $\mathcal{L}^{(4)}_{14}$ &$R_{abcd}R^{aecf}R_{gehf}R^{gbhd}$ &$8(n-1)(n^7-4n^6-15n^5+122n^4-287n^3+297n^2-126n+18)$\\
  \hline
\end{tabular}
\caption{The terms of $\mathcal{L}^{(4)}_{i}$ and their
coefficients.} \label{tab:quart}
\end{center}
\end{table}
\pagebreak

In Eq. (\ref{quintic}), the coefficients $c_{i}$ are:
\begin{center}
	\begin{longtable}{|c|c|c|}
		\caption{The terms of $\mathcal{L}^{(5)}_{i}$ and their
			coefficients.}\label{tab:quint}\\
		\hline
		\textbf{Lable} & \textbf{Term} &  $ c_i $ \\
		\hline\hline
		\endfirsthead
		\multicolumn{3}{c}%
		{\tablename\ \thetable\ -- \textit{Continued from previous page}} \\
		\hline
		\textbf{Lable} & \textbf{Term} & $ b $ \\
		\hline
		\endhead
		\hline \multicolumn{3}{r}{\textit{Continued on next page}} \\
		\endfoot
		\hline
		\endlastfoot
$\mathcal{L}^{(5)}_{1}$ & $R R_{b}^{a} R_{c}^{b} R_{d}^{c} R_{a}^{d}$ & $ \begin{array} {lcl}
	\frac{1}{n-2}( 22\,{n}^{12}+98\,{n}^{11}-4227\,{n}^{10}+26488\,{n}^{9}-34298\,{n}^{8} \\ -314764\,{n}^{7}+1879963\,{n}^{6}-5179230\,{n}^{5}+8667296\,{n}^{4} \\ -9278000\,{n}^{3}+6209228\,{n}^{2}-2352032\,n+379200)
\end{array} $
\\
  \hline
 $\mathcal{L}^{(5)}_{2}$ & $ R R_{b}^{a} R_{a}^{b} R_{ef}^{cd} R_{cd}^{ef}$ & $ \begin{array} {lcl} 9\,{n}^{11}+34\,{n}^{10}-1541\,{n}^{9}+11499\,{n}^{8}-25758
\,{n}^{7}-81964\,{n}^{6} \\ +660233\,{n}^{5} -1886059\,{n}^{4}+3046869\,{n}
^{3}-2977682\,{n}^{2} \\ +1666312\,n-41192 \end{array} $ \\
  \hline
  $\mathcal{L}^{(5)}_{3}$  & $R R_{c}^{a} R_{d}^{b} R_{ef}^{cd} R_{ab}^{ef}$ & $ \begin{array} {lcl} \frac{1}{2(n-2)}( -58\,{n}^{12}-162\,{n}^{11}+10663\,{n}^{10}-84812\,{
    n}^{9} \\ +229322\,{n}^{8}+436556\,{n}^{7}-5176607\,{n}^{6}+18005330\,{n}^
{5} \\ -35943244\,{n}^{4}+45563680\,{n}^{3} -36695932\,{n}^{2}+17330208\,n \\ -
3674560) \end{array} $ \\
 \hline
  $\mathcal{L}^{(5)}_{4}$ & $R_{b}^{a} R_{a}^{b} R_{d}^{c}  R_{e}^{d} R_{c}^{e}$ & $ \begin{array} {lcl} -\frac{2(n-1)}{n-2}( 9\,{n}^{11}+34\,{n}^{10}-1541\,{n}^{9}+11499\,{n}^{8}-25758\,{n}^{7} \\ -81964\,{n}^{6} +660233
\,{n}^{5}-1886059\,{n}^{4}+3046869\,{n}^{3} \\ -2977682\,{n}^{2}+1666312\,n  -411920) \end{array} $\\
 \hline
 $\mathcal{L}^{(5)}_{5}$ & $R_{b}^{a} R_{c}^{b} R_{a}^{c}  R_{fg}^{de} R_{de}^{fg}$ & $ \begin{array} {lcl} \frac{1}{4(n-2)}(208\,{n}^{13}-4737\,{n}^{12}+40968\,{n}^{11}-159932\,{n}^{10} \\ +101251\,{n}^{9} +1850607\,{n}^{8}-9772230\,{n}^{7}+27253898
\,{n}^{6} \\ -50334197\,{n}^{5} +65342916\,{n}^{4}-60349728\,{n}^{3}+
38913248\,{n}^{2} \\ -16207200\,n +3316864) \end{array} $ \\
 \hline
$\mathcal{L}^{(5)}_{6}$ & $ R_{b}^{a} R_{d}^{b} R_{f}^{c}  R_{ag}^{de} R_{ce}^{fg}$ & $ \begin{array} {lcl} \frac{1}{n-2}(-296\,{n}^{13}+5380\,{n}^{12}-47491\,{n}^{11}+235224\,{n}^{10} \\ -501416\,{n}^{9} -1195535\,{n}^{8}+12548311\,{n}^{7}-45635482
\,{n}^{6} \\  +100350946\,{n}^{5}  -146329207\,{n}^{4}+143219210\,{n}^{3} \\ -
91132732\,{n}^{2}+34380784\,n-5893664) \end{array} $\\
  \hline
 $\mathcal{L}^{(5)}_{7}$ & $R_{b}^{a} R_{d}^{b} R_{f}^{c} R_{cg}^{de} R_{ae}^{fg}$ & $ \begin{array} {lcl} \frac{1}{4(n-2)}(184\,{n}^{14}-3251\,{n}^{13}+28056\,{n}^{12}-109604\,{n}^{11} \\ +6501\,{n}^{10}+1605461\,{n}^{9}-5747494\,{n}^{8}+3380818\,{
    n}^{7} \\ +36068661\,{n}^{6} -144617644\,{n}^{5}+293846956\,{n}^{4} \\ -
373665156\,{n}^{3}+300154032\,{n}^{2} -139373056\,n \\ +28457216) \end{array} $\\
  \hline
 $\mathcal{L}^{(5)}_{8}$ & $R_{b}^{a} R_{c}^{b} R_{ae}^{cd} R_{gh}^{ef} R_{df}^{gh}$ & $ \begin{array} {lcl} \frac{1}{4(n-2)}(596\,{n}^{13}-7977\,{n}^{12}+71966\,{n}^{11}-505962\,{n}^{10} \\ +2089493\,{n}^{9} -2365377\,{n}^{8}-20061508\,{n}^{7}+
119539756\,{n}^{6} \\ -335696499\,{n}^{5} +581268584\,{n}^{4}-651267392\,{n
}^{3} \\ +461836336\,{n}^{2}-188752160\,n +33860352) \end{array} $\\
  \hline
 $\mathcal{L}^{(5)}_{9}$ & $R_{b}^{a} R_{c}^{b} R_{ef}^{cd} R_{gh}^{ef} R_{ad}^{gh}$ & $ \begin{array} {lcl} \frac{1}{16(n-2)}(-184\,{n}^{15}+3155\,{n}^{14}-31372\,{n}^{13}+214234\,{n}^{12} \\ -1011489\,{n}^{11}+2804783\,{n}^{10}+374252\,{n}^{9}-44192768\,{n}^{8} \\ +224431715\,{n}^{7}-655954220\,{n}^{6}+1293485398\,{n
}^{5} \\ -1792474880\,{n}^{4} +1739485312\,{n}^{3}-1131595440\,{n}^{2} \\ +
442875968\,n-78459392) \end{array} $\\
  \hline
 $\mathcal{L}^{(5)}_{10}$ & $R_{b}^{a} R_{c}^{b} R_{eg}^{cd} R_{ah}^{ef} R_{df}^{gh}$ & $ \begin{array} {lcl} \frac{1}{2(n-2)}(304\,{n}^{14}-5487\,{n}^{13}+51364\,{n}^{12}-296956\,{n}^{11} \\ +1020583\,{n}^{10} -1134859\,{n}^{9}-8135394\,{n}^{8}+
52879112\,{n}^{7} \\ -168012561\,{n}^{6} +347472004\,{n}^{5}-498259688\,{n}
^{4} \\ +497441450\,{n}^{3}-331820224\,{n}^{2} +132631584\,n \\ -23851392) \end{array} $\\
  \hline
 $\mathcal{L}^{(5)}_{11}$ & $R_{c}^{a} R_{d}^{b} R_{ab}^{cd} R_{gh}^{ef} R_{ef}^{gh}$ & $ \begin{array} {lcl} \frac{1}{4(n-2)}(-244\,{n}^{13}+4709\,{n}^{12}-34468\,{n}^{11}+95172\,{n}^{10} \\ +152097\,{n}^{9} -1923839\,{n}^{8}+6353794\,{n}^{7}-11131154
\,{n}^{6} \\ +10232149\,{n}^{5} -1781288\,{n}^{4}-6422656\,{n}^{3}+6551632
\,{n}^{2} \\ -2066336\,n-21504) \end{array} $\\
  \hline
 $\mathcal{L}^{(5)}_{12}$ & $R_{c}^{a} R_{d}^{b} R_{ae}^{cd} R_{gh}^{ef} R_{bf}^{gh}$ & $ \begin{array} {lcl} \frac{1}{8(n-2)}(416\,{n}^{14}-10647\,{n}^{13}+86586\,{n}^{12} \\ -223848\,{n}^{11}-764407\,{n}^{10} +6904499\,{n}^{9}-18735836\,{n}^{8} \\ +
11482750\,{n}^{7}+69049061\,{n}^{6} -239246282\,{n}^{5} \\ +400589060\,{n}^
{4}-410760584\,{n}^{3}+261506352\,{n}^{2} \\ -94377920\,n+14506368) \end{array} $\\
  \hline
 $\mathcal{L}^{(5)}_{13}$ & $R_{c}^{a} R_{d}^{b} R_{ef}^{cd} R_{gh}^{ef} R_{ab}^{gh}$ & $ \begin{array} {lcl} \frac{1}{16(n-2)}(-184\,{n}^{15}+4003\,{n}^{14}-34770\,{n}^{13}+206558\,{n}^{12} \\ -1209685\,{n}^{11}+6001605\,{n}^{10}-16647870\,{n}^{9} \\ 
-3841080\,{n}^{8}+218943659\,{n}^{7}-902806270\,{n}^{6} \\ +2083343490\,{n
}^{5}-3136302944\,{n}^{4}+3171015856\,{n}^{3} \\ -2094129968\,{n}^{2}+
819673024\,n-144243200) \end{array}$\\
  \hline
 $\mathcal{L}^{(5)}_{14}$ & $R_{c}^{a} R_{d}^{b} R_{eg}^{cd} R_{ah}^{ef} R_{bf}^{gh}$ & $ \begin{array} {lcl} \frac{1}{2(n-2)}(388\,{n}^{14}-4716\,{n}^{13}+29243\,{n}^{12}-136746\,{n}^{11} \\ +450540\,{n}^{10}-132929\,{n}^{9}-8134503\,{n}^{8}+48332850
\,{n}^{7} \\ -155977854\,{n}^{6}+329810835\,{n}^{5}-478872342\,{n}^{4} \\ +
476176930\,{n}^{3}-310557920\,{n}^{2}+119520320\,n-20516736) \end{array}$\\
  \hline
 $\mathcal{L}^{(5)}_{15}$ & $R_{c}^{a} R_{e}^{b} R_{af}^{cd} R_{gh}^{ef} R_{bd}^{gh}$ & $ \begin{array} {lcl} \frac{1}{8(n-2)}(664\,{n}^{14}-9139\,{n}^{13}+57128\,{n}^{12}-185108\,{n}^{11} \\ +159381\,{n}^{10}+1223517\,{n}^{9}-6100234\,{n}^{8}+14465638
\,{n}^{7} \\ -20079091\,{n}^{6}+16176660\,{n}^{5}-10250920\,{n}^{4} \\ +
14943120\,{n}^{3}-20032144\,{n}^{2}+11852416\,n-2285952) \end{array}$\\
  \hline
 $\mathcal{L}^{(5)}_{16}$ & $R_{b}^{a} R_{ad}^{bc} R_{fh}^{de} R_{ci}^{fg} R_{eg}^{hi}$ & $ \begin{array} {lcl} \frac{1}{2}(540\,{n}^{13}-10293\,{n}^{12}+81315\,{n}^{11}-291890\,{n}^{10} \\ +61415\,
{n}^{9}+3807202\,{n}^{8}-16976001\,{n}^{7}+38237858\,{n}^{6} \\ -48573651
\,{n}^{5}+26466351\,{n}^{4}+15817758\,{n}^{3}-37469132\,{n}^{2} \\ +
25084592\,n-6246112) \end{array}$\\
 \hline
 $\mathcal{L}^{(5)}_{17}$ & $R_{b}^{a} R_{de}^{bc} R_{cf}^{de} R_{hi}^{fg} R_{ag}^{hi}$ & $ \begin{array} {lcl} \frac{1}{16(n-2)}(432\,{n}^{15}-4127\,{n}^{14}+19469\,{n}^{13}-170554\,{n}^{12} \\ +1675605\,{n}^{11}-9507738\,{n}^{10}+29711901\,{n}^{9}-
40384306\,{n}^{8} \\ -54483045\,{n}^{7}+383370077\,{n}^{6}-912069066\,{n}^
{5} \\ +1331612328\,{n}^{4}-1311989568\,{n}^{3}+870354912\,{n}^{2} \\ -
355966464\,n+67907584) \end{array} $\\
  \hline
 $\mathcal{L}^{(5)}_{18}$ & $R_{b}^{a} R_{df}^{bc} R_{ac}^{de} R_{hi}^{fg} R_{eg}^{hi}$ & $ \begin{array} {lcl} \frac{1}{2(n-2)}(62\,{n}^{15}-261\,{n}^{14}+82\,{n}^{13}-34985\,{n}^{12}+465930\,{n}^{11} \\ -2557591\,{n}^{10}+6958394\,{n}^{9}-5370935\,{n}^{8}-27811996\,{n}^{7} \\ +116102040\,{n}^{6}-231876220\,{n}^{5}+291631996
\,{n}^{4} \\ -242759516\,{n}^{3}+131704680\,{n}^{2}-43058976\,n+6606272) \end{array} $\\
  \hline
 $\mathcal{L}^{(5)}_{19}$ & $ R_{b}^{a} R_{df}^{bc} R_{ah}^{de} R_{ei}^{fg} R_{cg}^{hi}$ & $ \begin{array} {lcl} \frac{1}{4(n-2)}(-656\,{n}^{15}+8832\,{n}^{14}-54341\,{n}^{13}+172912\,{n}^{12} \\ -46160\,{n}^{11}-2326941\,{n}^{10}+11290819\,{n}^{9}-
23788482\,{n}^{8} \\ +3603422\,{n}^{7}+115748491\,{n}^{6}-336672132\,{n}^{
    5} \\ +513849092\,{n}^{4}-485503192\,{n}^{3}+287262016\,{n}^{2} \\ -98692160\,
n+15146752) \end{array}$\\
  \hline
 $\mathcal{L}^{(5)}_{20}$ & $R_{b}^{a} R_{df}^{bc} R_{gh}^{de} R_{ei}^{fg} R_{ac}^{hi}$ & $ \begin{array} {lcl} \frac{1}{8(n-2)}(-1328\,{n}^{15}+24603\,{n}^{14}-201582\,{n}^{13}+847816\,{n}^{12} \\ -1334949\,{n}^{11}-4313683\,{n}^{10}+32443416\,{n}^{9} \\ -106895622\,{n}^{8}+248652911\,{n}^{7}-456679514\,{n}^{6} \\ +663949044\,{
    n}^{5}-732715856\,{n}^{4}+587021544\,{n}^{3} \\ -326614080\,{n}^{2}+
116003840\,n-20182272) \end{array}$\\
 \hline
 $\mathcal{L}^{(5)}_{21}$ & $R_{cd}^{ab} R_{eg}^{cd} R_{ai}^{ef} R_{fj}^{gh}R_{bh}^{ij}$ & $ \begin{array} {lcl} \frac{1}{8(n-2)}(184\,{n}^{16}-3339\,{n}^{15}+27760\,{n}^{14}-131212\,{n}^{13} \\ +355561\,{n}^{12}-357883\,{n}^{11}-1572146\,{n}^{10}+
11274022\,{n}^{9} \\ -44976839\,{n}^{8}+128367156\,{n}^{7}-256851408\,{n}^
{6} \\ +335947624\,{n}^{5}-240332008\,{n}^{4}+11641104\,{n}^{3} \\ +134199392
\,{n}^{2}-104249344\,n+26702336) \end{array} $\\
 \hline
 $\mathcal{L}^{(5)}_{22}$ & $R_{ce}^{ab} R_{af}^{cd} R_{gi}^{ef} R_{bj}^{gh}R_{dh}^{ij}$ & $ \begin{array} {lcl} \frac{1}{2(n-2)}(-284\,{n}^{15}+4973\,{n}^{14}-37942\,{n}^{13}+144773\,{n}^{12} \\ -109479\,{n}^{11}-1875825\,{n}^{10}
+12234317\,{n}^{9}-
45166705\,{n}^{8} \\ +119677671\,{n}^{7}-240530864\,{n}^{6}+367236029\,{n}
^{5} \\ -416310288\,{n}^{4}+337180200\,{n}^{3}-183807888\,{n}^{2} \\ +60487840
\,n-9120896) \end{array}$\\
 \hline
 $\mathcal{L}^{(5)}_{23}$ & $R_{ce}^{ab} R_{ag}^{cd} R_{bi}^{ef} R_{fj}^{gh}R_{dh}^{ij}$ & $ \begin{array} {lcl} \frac{1}{4(n-2)}(-8\,{n}^{15}+2019\,{n}^{14}-38926\,{n}^{13}+337600\,{n}^{12} \\ -1605181\,{n}^{11}+3785705\,{n}^{10}+1659444\,{n}^{9} \\ -
45775086\,{n}^{8}+176674471\,{n}^{7}-400284742\,{n}^{6} \\ +611871600\,{n}
^{5}-648934536\,{n}^{4}+469233704\,{n}^{3} \\ -218274992\,{n}^{2}+57719936\,n-6343424) \end{array}$\\
  \hline
 $\mathcal{L}^{(5)}_{24}$ &$R_{ce}^{ab} R_{fg}^{cd} R_{hi}^{ef} R_{aj}^{gh}R_{bd}^{ij}$ &$ \begin{array} {lcl} \frac{1}{4}(184\,{n}^{15}-3179\,{n}^{14}+25777\,{n}^{13}-115454\,{n}^{12} \\ +228481\,
{n}^{11}+522238\,{n}^{10}-5783003\,{n}^{9}+23848974\,{n}^{8} \\ -64717433
\,{n}^{7}+128477225\,{n}^{6}-193789406\,{n}^{5} \\ +224224860\,{n}^{4}-
195140632\,{n}^{3}+120313912\,{n}^{2} \\ -46440128\,n+8345216) \end{array}$\\
  \hline
\end{longtable}
\end{center}


\end{document}